\documentclass[useAMS,usenatbib]{mn2e}

\usepackage{graphicx}
\usepackage{amsmath}
\usepackage{amssymb}
\usepackage{color}
\usepackage{gensymb}
\usepackage{subfigure}

\title[HVSs from young stellar clusters in the GC]{Hypervelocity stars from young stellar clusters in the Galactic Centre}
\author[G. Fragione, R. Capuzzo-Dolcetta and P. Kroupa]{G. Fragione$^{1}$\thanks{E-mail: giacomo.fragione@uniroma1.it}, R. Capuzzo-Dolcetta$^{1}$ and P. Kroupa$^{2,3}$\\
$^{1}$Department of Physics, Sapienza, University of Rome, P.le A. Moro 2, 00185 Roma, Italy\\
$^{2}$Helmholtz-Institut f\"{u}r Strahlen-und Kernphysik, Universit\"{a}t Bonn, Nussallee 14-16, 53115 Bonn, Germany\\
$^{3}$Charles University in Prague, Faculty of Mathematics and Physics, Astronomical Institute, V  Hole\v{s}ovi\v{c}k\'ach 2,\\CZ-180 00 Praha 8, Czech Republic}
\begin{document}

\maketitle

\begin{abstract}
The enormous velocities of the so called hypervelocity stars (HVSs) derive, likely, from close interactions with massive black holes, binary stars encounters or supernova explosions. In this paper, we investigate the origin of hypervelocity stars as consequence of the close interaction between the Milky Way central massive black hole and a passing-by young stellar cluster. We found that both single and binary HVSs may be generated in a burst-like event, as the cluster passes near the orbital pericentre. High velocity stars will move close to the initial cluster orbital plane and in the direction of the cluster orbital motion at the pericentre. The binary fraction of these HVS jets depends on the primordial binary fraction in the young cluster. The level of initial mass segregation determines the value of the average mass of the ejected stars. Some binary stars will merge, continuing their travel across and out of the Galaxy as blue stragglers.
\end{abstract}


\label{firstpage}

\section{Introduction}
\label{sec:intro}

Hypervelocity Stars (HVSs) are stars escaping the Milky Way (MW) gravitational well. Whereas \citet{hil88} predicted theoretically their existence, HVSs were observed for the first time only $17$ years later by \citet{brw05}. About $20$ HVSs have been found at velocities up to $\approx 700$ km s$^{-1}$ in the outer MW halo (between $50$ and $120$ kpc) by the Multiple Mirror Telescope (MMT) spectroscopic survey \citep{brw10,brw14}. This survey targets stars with magnitudes and colors typical of $2.5-4\ \mathrm{M}_{\odot}$ late B-type stars. As consequence of the target strategy, the MMT stars could be either main sequence B stars, evolved blue horizontal branch stars or blue stragglers \citep{brw14}. Recently, astronomers have started to investigate low-mass HVS candidates \citep{li15,zie15}. Being a MMT spectroscopic survey, HVS data suffer from the lack of tangential velocity measurements. The astrometric European satellite \textit{Gaia} (http://www.cosmos.esa.int/web/gaia) is expected to measure proper motions with an unprecedented precision, providing a larger and less biased sample of HVSs (allegedly $\approx 100$ in a catalogue of $\approx 10^9$ stars). Moreover, \textit{Gaia}'s sensitivity is good enough to search for multi-planet systems around massive stars and to reveal their architectures and three-dimensional orbits, giving the possibility of spotting some planetary transits around HVSs \citep*{gin12,frg16}.

The Hills' mechanism \citep{hil88} involves the tidal breakup of a binary that passes close to the Milky Way Black Hole (BH) \citep*{gil06,gil07,loc08,oll08,sar09}, but the physical mechanism responsible for the production of the observed HVSs is still debated. Other possible origins have been advanced \citep{brw15}, such as the interaction of a BH binary with a single star \citep{yut03}, the interaction of star clusters and BHs \citep{cap15,fra16}, supernova explosions \citep*{zub13,tau15}, tidal disruption of a dwarf galaxy passing through the Galactic Center (GC) \citep{aba09} and the dynamical evolution of a thin and eccentric disk orbiting around a massive BH \citep{sub14,haa16,sub16}. The extreme velocities of HVSs indicate they may derive from a strong dynamical interaction with BHs in the GC or in a nearby galaxy \citep*{she08}.

The importance of HVSs is that they can provide information on the environment where they were born \citep{gou03}. In particular, they can discriminate between a single and a binary BH in the GC \citep*{ses07}. Since their orbits are determined by the MW potential, HVS kinematics can be used to probe the Galactic potential's triaxiality \citep*{gne05,yum07} and discriminate among different Galactic mass distributions \citep*{pee09,gne10,frl16}.

As \citet{cap15} and \citet{fra16} have shown, a relevant mechanism to accelerate stars to high or even hyper velocities is that due to the close interaction of a single or binary massive black hole and a passing-by massive stellar cluster \citep*{arc16}. In the cited works the analysis was done for an evolved globular cluster-like object. The aim of this paper is, instead, that of investigating a possible origin of HVSs which involves a Young Star Cluster (YSC) that, during its orbit, has had the chance to pass close to the MW central super massive black hole (SMBH). When the YSC passes by the SMBH, some of its stars can be stripped from the cluster and ejected with high velocities \citep{cap15,fra16}.

The outline of the paper is as follows. In Section 2, we describe the method we use to study the consequences of YSC-BH interaction. In Section 3, the results are presented and discussed. Finally, in Section 4, we summarize our main conclusions.

\section{Method}
\label{sec:meth}

We use the publicly available code \textsc{McLuster} to generate initial conditions for the YSC \citep{kup11}. \textsc{McLuster} sets up initial conditions once the desired features of the cluster are specified, such as the total mass $M_{cl}$ (or alternatively the total number of stars $N$), the density profile, the initial degree of mass segregation $S$ and the primordial binary fraction $B$.

We studied a YSC interacting with the SgrA* SMBH via direct $N$-body simulations. To do this, we consider a set of $6$ cluster models, which differ in the initial binary fraction ($B$) and in the initial degree of mass segregation ($S$). $B$ is defined as
\begin{equation}
B=\frac{N_b}{N_b+N_s},
\end{equation}
where $N_b$ is the number of binaries and $N_s$ is the number of single stars in the cluster. The initial binary fraction of our models assumes the values $B=0,0.25,0.5,0.75,1$ for both segregated ($S=1$) and unsegregated ($S=0$) clusters. Note that some of our models have a high initial binary fraction. Such an assumption is justified by the fact that both theory and observations suggest that the angular momentum of a collapsing cloud core may be distributed more efficiently into two stars \citep{goo05,mar12,lei15,ohk15} rather than into a single star or a higher order multipole system. Initially mass-segregated clusters are generated with the method developed in \citet*{bau08}. In a few words: if the cluster is completely mass segregated ($S=1$), \textsc{McLuster} assigns the lowest energy orbit to the highest mass star, and the highest energy orbit to the lowest mass star. Intermediate degrees of mass segregation can be achieved by non-perfect ordering of masses and energy orbits \citep{kup11}. Table 1 resumes the cases considered.

To generate initial conditions, we follow the prescriptions of \citet*{ohk15}. We set the mass of the cluster to $M_{cl}=10^3$ M$_{\odot}$. We note here that $10^4$ M$_{\odot}$ clusters are observed more easily than lighter ones due to an observational bias. Star formation in the inner region of the Galaxy is likely to be producing many more low-mass clusters, which are hardly observable given the dust absorption in the whole inner Galactic region. This is why we are studying a, likely, common very young cluster, to understand whether such a cluster may lead to an anisotropic flux of HVSs. The cluster density distribution is modelled with a \citet{plu11} profile whose
initial half-mass radius is derived from the \citet{mar12} relation
\begin{equation}
r_h=0.1 \times (M_{cl}/\mathrm{M}_{\odot})^{0.13}\ \mathrm{pc}=0.25\ \mathrm{pc}.
\end{equation}

The stellar masses, $m$ (M$_\odot$), are sampled from the initial mass function (IMF) by \citet*{kro01,kro13}:
\begin{equation}
\xi(m)=
\begin{cases}
k_1\left(\frac{m}{0.08}\right)^{-1.3}& \text{$m_{min}\le m/\mathrm{M}_\odot\leq 0.50$},\\
k_2\left(\frac{0.5}{0.08}\right)^{-1.3}\left(\frac{m}{0.5}\right)^{-2.3}& \text{$0.50\le m/\mathrm{M}_\odot\leq m_{max}$},
\end{cases}
\label{eqn:imf}
\end{equation}
where $k_1\approx 0.113$ and $k_2\approx 0.035$ are normalization factors, $m_{min}=0.08$ M$_\odot$, and $m_{max}=40.54$ M$_\odot$ is chosen from the maximum stellar mass-cluster mass relation given in \citet{wei04} and \citet{pfl07}. 

The canonical \citet{kro95b} period distribution function is adopted for all binaries in the cluster
\begin{equation}
f(\log_{10}P)=2.5\frac{\log_{10}P-1}{45+(\log_{10}P-1)^2},
\end{equation}
where the period, $P$, is in days. Stars with mass $m<5$ M$_{\odot}$ are randomly paired after sampling the two masses from the chosen IMF \citep{kro95a,kro95b}. However, observations show that massive binaries preferentially favour massive companions \citep{san12}, therefore, stars with $m\ge 5$ M$_{\odot}$ are paired together to follow this observational feature; for details see \citet{ohk15}. The initial binary eccentricities, $e$, are drawn from a thermal distribution  \citep{kro08},
\begin{equation}
f_e(e)=2e.
\end{equation}

\begin{table}
\caption{Cluster models: name, initial binary fraction ($B$) and initial mass segregation fraction ($S$).}
\centering
\begin{tabular}{l|c|c|c}
\hline
Name & $B$ & $S$\\
\hline
B1S1  & 1   & 1\\
B1S0  & 1   & 0\\
B0S1  & 0   & 1\\
B025S1 & 0.25 & 1\\
B05S1 & 0.5 & 1\\
B075S1 & 0.75 & 1\\
\hline
\end{tabular}
\label{tab1}
\end{table}

All the models are generated under the assumption that the clusters are in virial equilibrium \citep{ohk15}. The time integration of the stellar system was done by means of the publicly available code \textsc{nbody6} \citep{aar03}. Initally, the cluster center-of-mass is set in the Galactic disk on the $x$-axis at $100$ pc from the GC, with an initial velocity of $2$ km s$^{-1}$ along the Galactic $y$-axis (the pericenter results to be $\approx 0.45$ pc and the orbital eccentricity $e_{cl}\approx 0.99$). The choice of a highly eccentric orbit is justified by our interest to study the maximum possible effect in the massive BH-YSC interaction that depends mainly on the distance of closest approach (for given values of other parameters). The role of changing the eccentricity, and so changing the pericentric distance, has been already investigated in \citet{cap15} and \citet{fra16}. In particular, \citet{cap15} showed that the larger the eccentricity the more efficient the ejection of high-velocity stars. This effect combines with the total mass of the cluster, where a more massive cluster generates a larger fraction of fast-moving stars (as also discussed in Sect. 3.1).

We produced $150$ different realizations of each model (generated with different random seeds) to obtain statistically relevant results.

\subsection{The Milky Way potential}

We describe the MW potential with a 4-component  model $\Phi(r)=\Phi_{BH}+\Phi_{b}(r)+\Phi_{d}(r)+\Phi_{h}(r)$ \citep{ken08,ken14,frl16}, where:

\begin{itemize}
\item $\Phi_{BH}$ is the contribution of the central SMBH,
\begin{equation}
\Phi_{BH}(r)=-\frac{GM_{BH}}{r},
\end{equation}
with mass $M_{BH}=4 \times 10^6$ M$_{\odot}$;
\item $\Phi_{b}$ is the contribution of the spherical bulge \citep{her90},
\begin{equation}
\Phi_{b}(r)=-\frac{GM_{bul}}{r+a},
\end{equation}
with mass $M_{b}=3.76\times 10^9$ M$_{\odot}$ and core radius $a=0.10$ kpc;
\item $\Phi_{d}$ accounts for the axisymmetric disc \citep{miy75},
\begin{equation}
\Phi_{disk}(R,z)=-\frac{GM_{disk}}{\sqrt(R^2+(b+\sqrt{c^2+z^2})^2)},
\end{equation}
with mass $M_{disk}=5.36\times 10^{10}$ M$_{\odot}$, length scale $b=2.75$ kpc and scale height $c=0.30$ kpc;
\item $\Phi_{halo}$ is the contribution of the dark matter halo \citep*{nav97}
\begin{equation}
\Phi_{halo}(r)=-\frac{GM_{DM}\ln(1+r/r_s)}{r}.
\end{equation}
with $M_{DM}=10^{12}$ M$_{\odot}$ and length scale $r_s=20$ kpc.
\end{itemize}

The parameters are chosen so that the Galactic circular velocity at the distance of the Sun ($8.15$ kpc) is $235$ km s$^{-1}$ \citep{rei14}.

\section{Results}
\label{sec:res}

\begin{figure}
\centering
\includegraphics[width=8.7cm,height=7.cm]{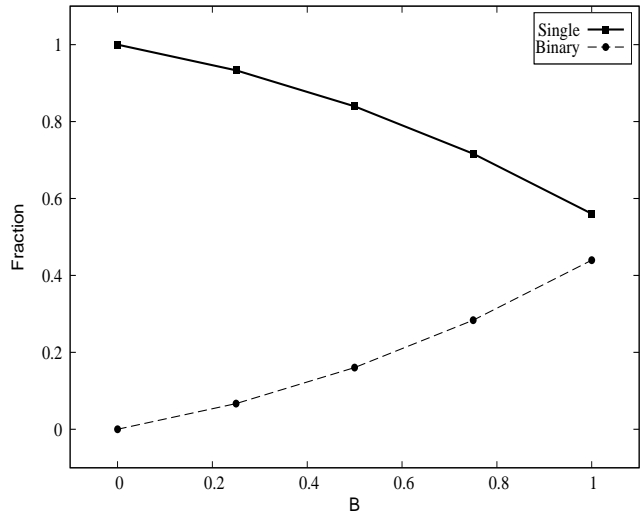}
\caption{Fraction of single ($N_{s,HVS}/N_{HVS}$) and binary ($N_{b,HVS}/N_{HVS}$) HVSs as function of the cluster binarity.}
\label{fractions}
\end{figure}

\begin{figure*}
\centering
\begin{minipage}{20.5cm}
\subfigure{\includegraphics[width=8.75cm,height=7.5cm]{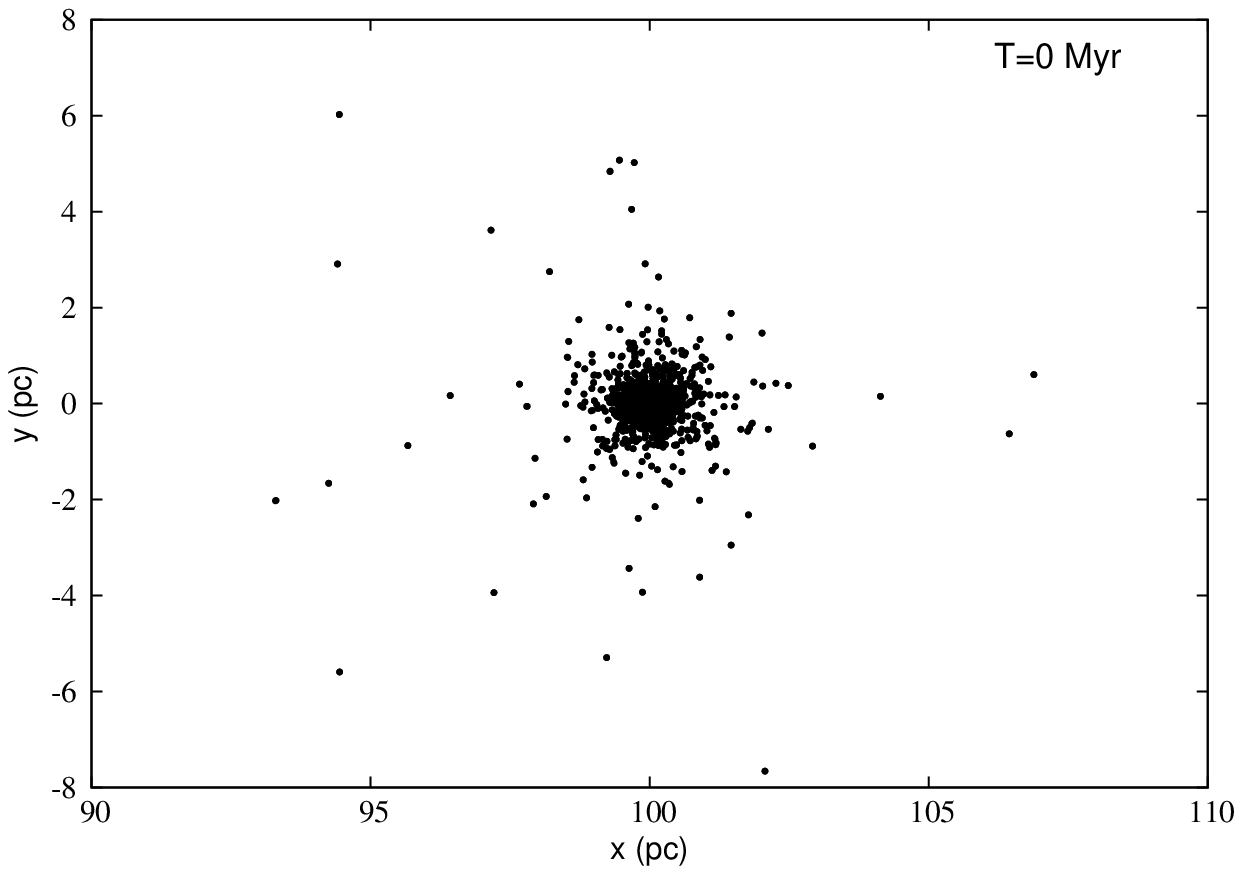}}
\subfigure{\includegraphics[width=8.75cm,height=7.5cm]{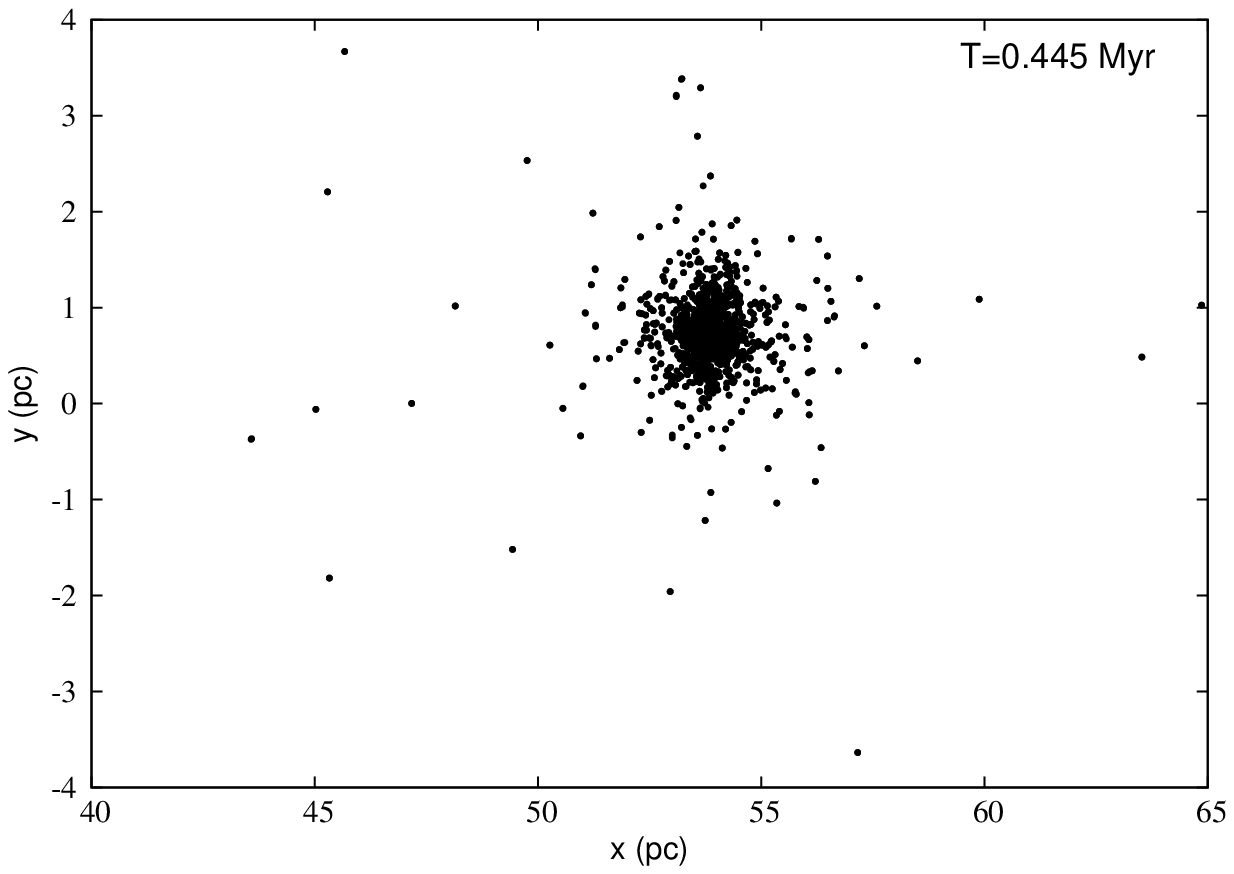}}
\end{minipage}
\begin{minipage}{20.5cm}
\subfigure{\includegraphics[width=8.75cm,height=7.5cm]{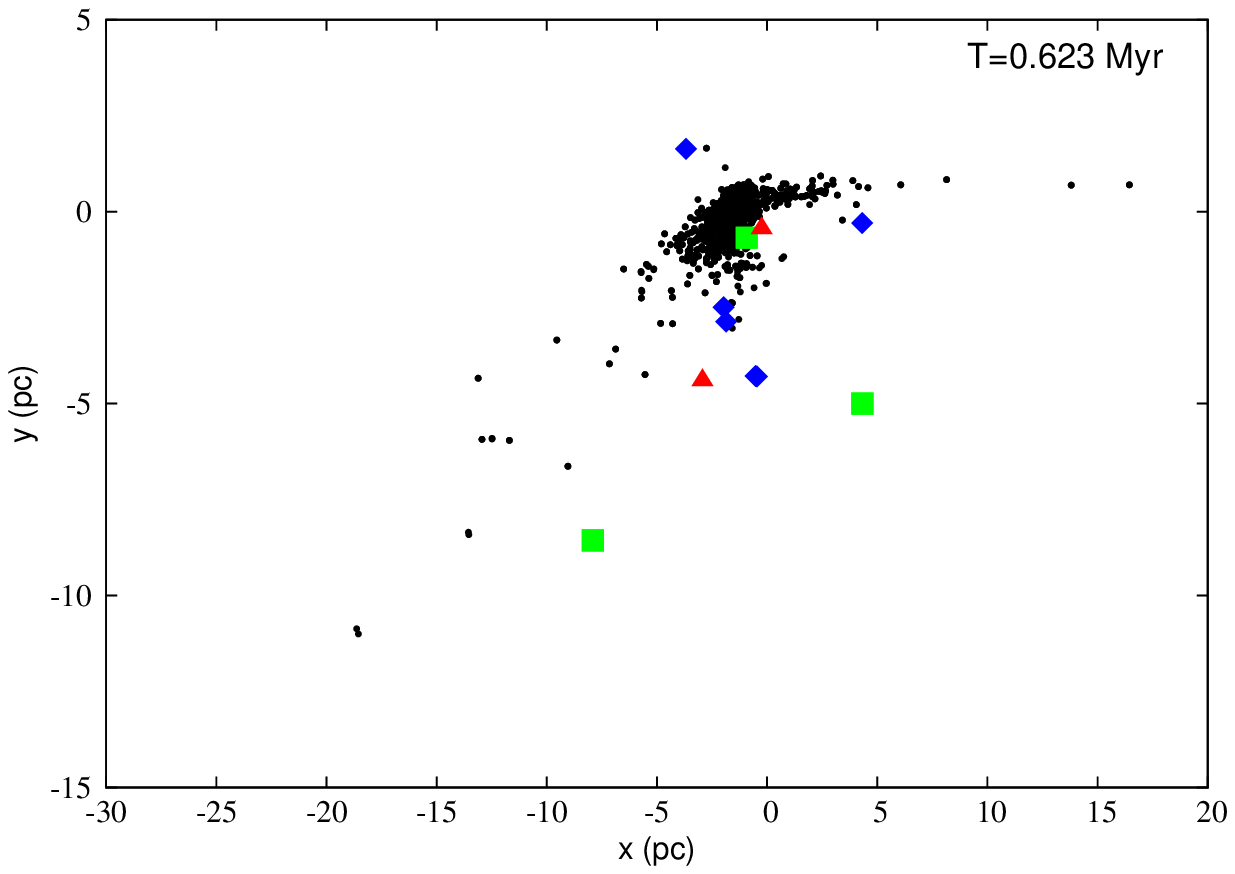}}
\subfigure{\includegraphics[width=8.75cm,height=7.5cm]{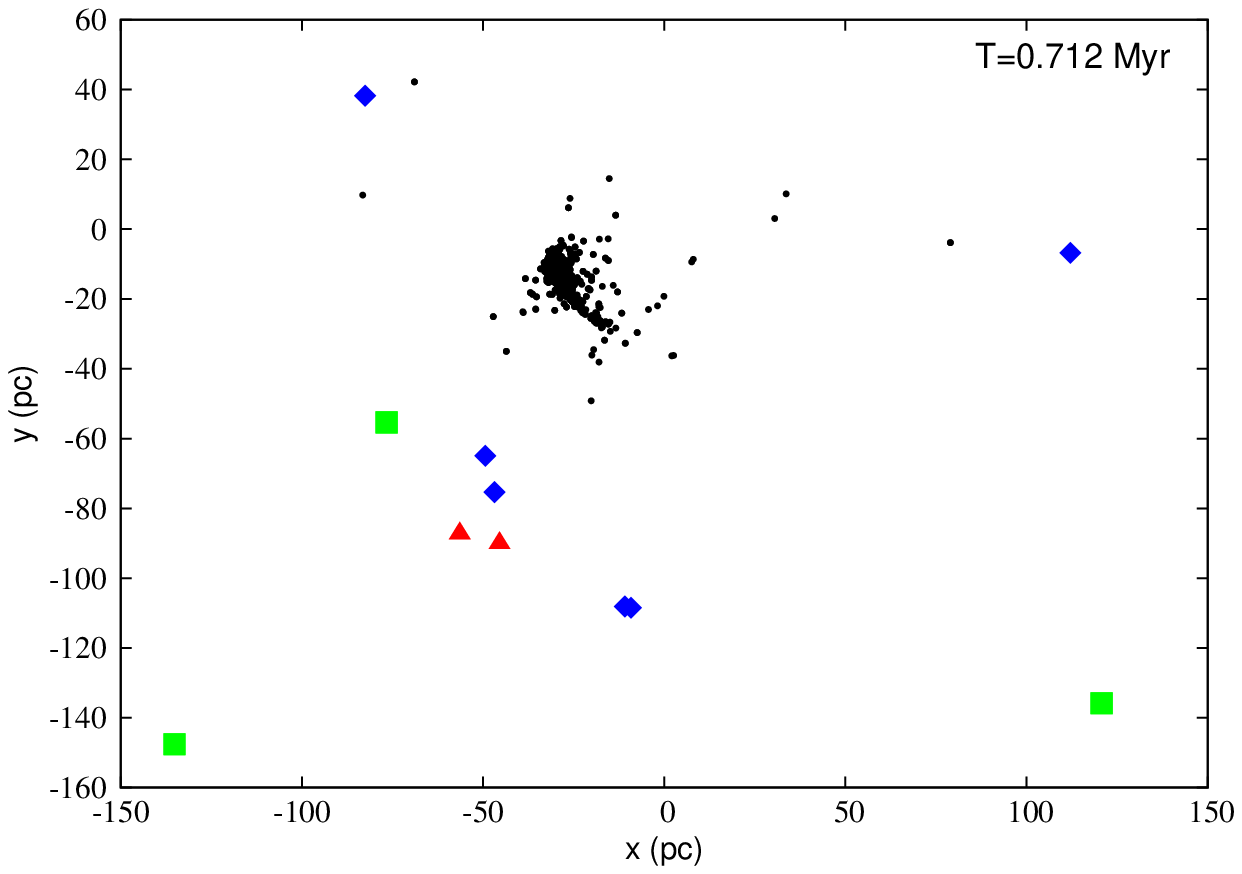}}
\end{minipage}
\caption{Snapshots at different times of the cluster model B1S1. After the pericentre passage, single and binary HVSs are produced. In the bottom panels, green squares represent binary HVSs, red triangles represent single HVS produced via the first two mechanisms described in Sect. 3, blue diamonds represent HVSs produced via the third mechanism described in Sect. 3. The massive BH has coordinates $(x,y,z)=(0,0,0)$.}
\label{snaps}
\end{figure*}

In our simulations the fate of the stars in the cluster is determined by their final amount of energy. After the close cluster-BH interaction a star can remain bound to the cluster \citep*{arc16}, or can be captured by the BH \citep*{gou03,gil06,per07} or can be lost from the system \citep{cap15,fra16}. In this paper, we focus our attention on the stars lost by the system and ejected at such velocities to become unbound with respect to the Galaxy, examining, in this context, the role of the cluster initial binary fraction and mass segregation. Figure \ref{snaps} shows snapshots at different times of the cluster model B1S1, and the production of single and binary HVSs after the close interaction with the BH.

The relative (to the total) number of ejected (single and binary) HVSs depends on the initial content of binaries (see Fig. \ref{fractions}). As $B$ increases, the number of ejected binary HVSs, $N_{b,HVS}$, increases. On the other hand, the relative fraction $N_{s,HVS}/N_{HVS}$, where $N_{s,HVS}$ and $N_{HVS}=N_{s,HVS}+N_{b,HVS}$ are the number of single HVSs and the total number of HVS systems, respectively, of ejected single HVSs decreases as the initial content of binaries in the cluster is larger. 

We find three possible sources of single HVSs. The first mechanism is equivalent to Hills' binary disruption \citep{hil88}. In this process HVS is generated as a consequence of the disruption of a binary star that undergoes a close interaction with the Galactic SMBH. In this case  one of the two stars is captured by the SMBH and becomes an S-star \citep{gou03,per07,loc08}. The second production channel involves a single star and the other $N-1$ stars of the cluster. During the close passage of the cluster around the SMBH, a single star may be removed from the cluster due to the coupled YSC+BH gravitational interaction \citep{cap15,fra16}. The third mechanisms is the splitting of a binary HVS into two single HVSs. Whereas single HVSs can originate from three different mechanisms, binary HVSs can originate only from the previous process. Actually, after the ejection from the cluster the binary can undergo three different fates. One possibility is that the binary star survives, travelling farther at hyper velocity. If the binary does not survive, it can either merge or disrupt as a consequence of the velocity kick due to the interaction with the SMBH, splitting into two single HVSs. This is, actually, the third channel for producing single HVSs. When the binary is accelerated, the kick velocity has to be combined with the binary orbital velocity. For simplicity, let us consider the case where the kick velocity vector is coplanar with the orbital velocity vector of the stars. It may happen that at the moment of the ejection the velocity of one of the stars in the pair is nearly aligned with the kick, while the companion velocity is unaligned. In this case, the orbital velocity of the former enhances the kick velocity, while the latter has a slightly lower ejection velocity. As a consequence, both the stars of the binary are accelerated to hyper velocities, but start to diverge from each other.

\begin{figure*}
\centering
\begin{minipage}{20.5cm}
\vspace{-0.5cm}
\subfigure{\includegraphics[width=8.75cm,height=7.5cm]{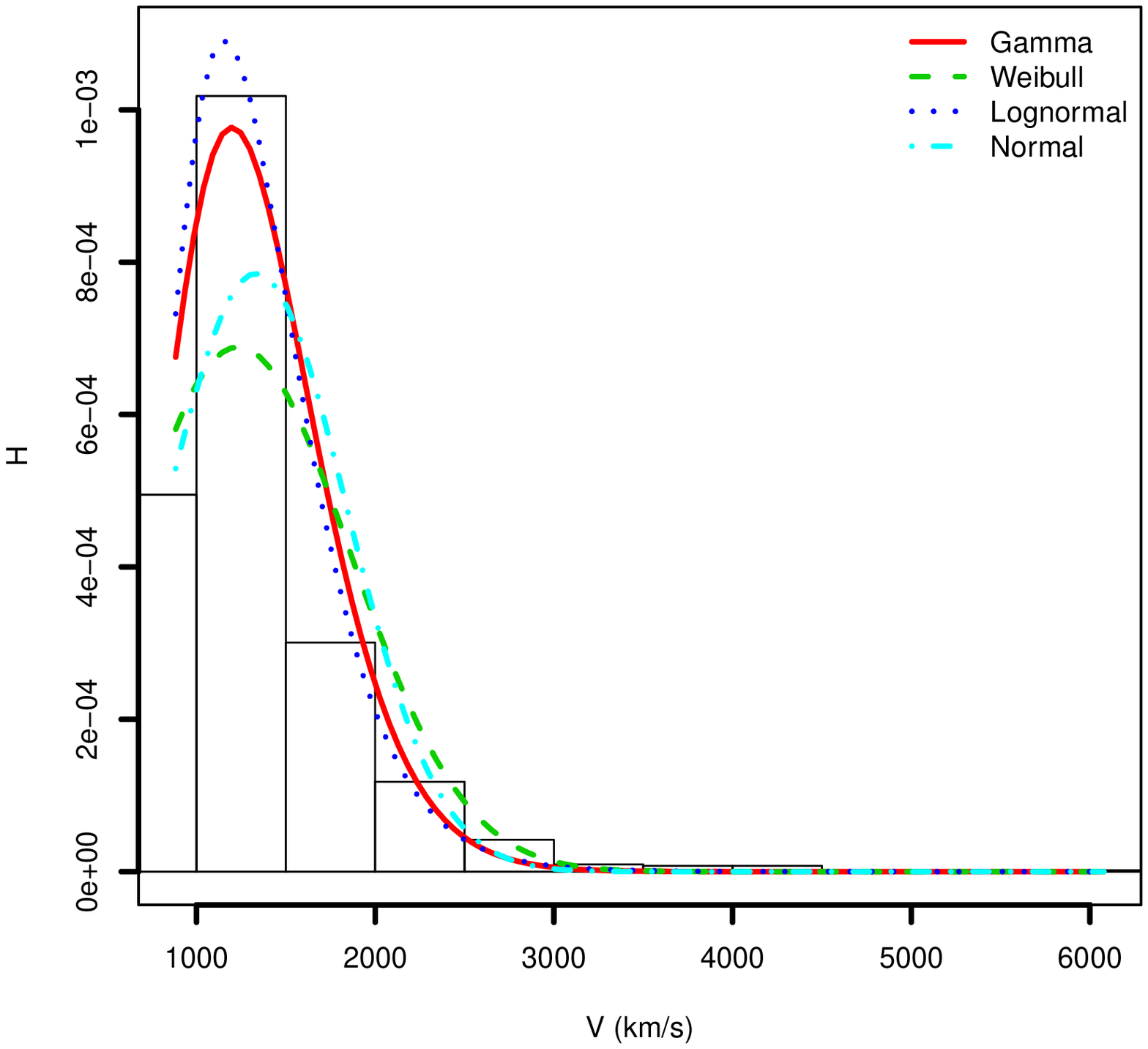}}
\subfigure{\includegraphics[width=8.75cm,height=7.5cm]{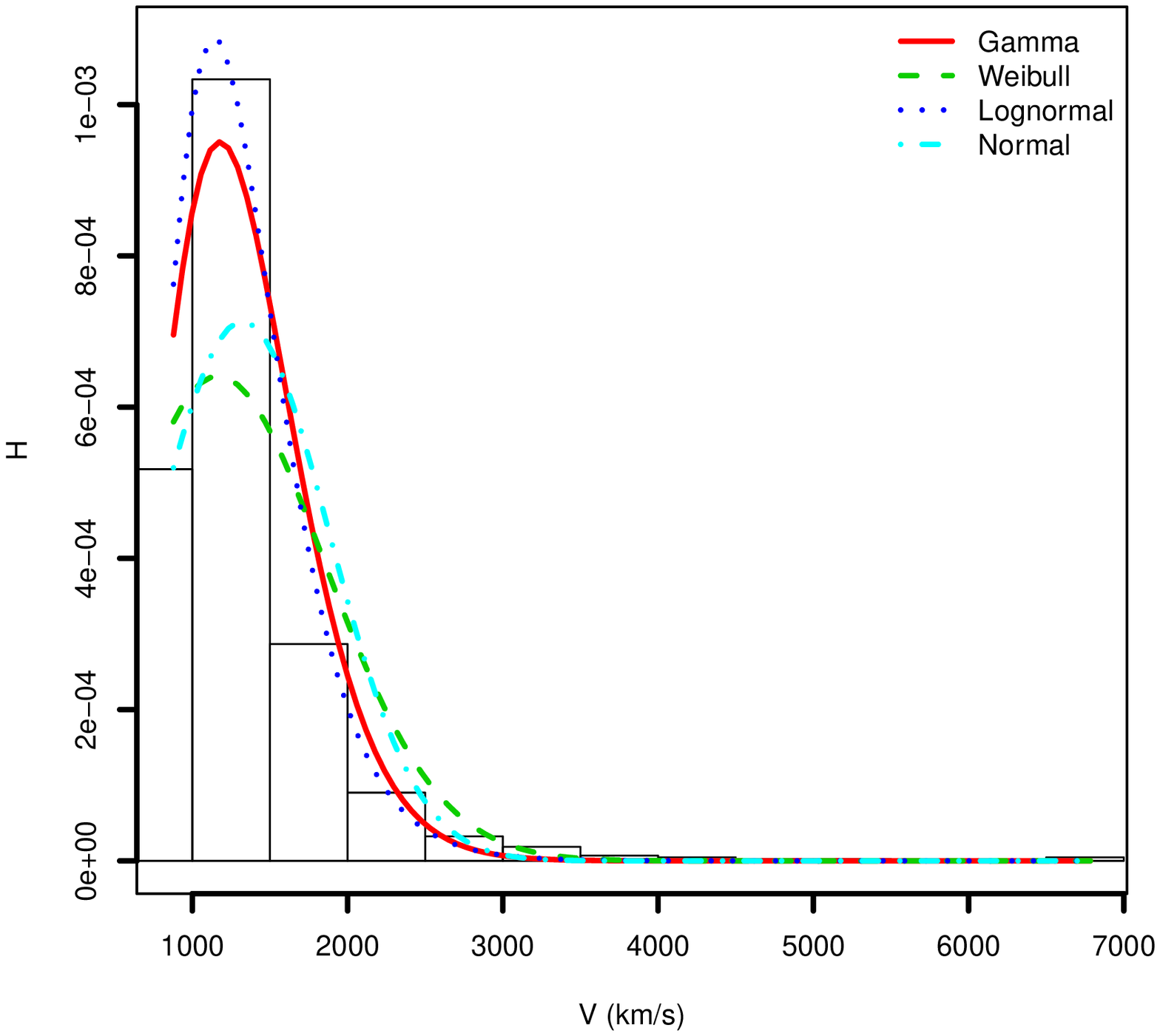}}
\end{minipage}
\begin{minipage}{20.5cm}
\vspace{-1cm}
\subfigure{\includegraphics[width=8.75cm,height=7.5cm]{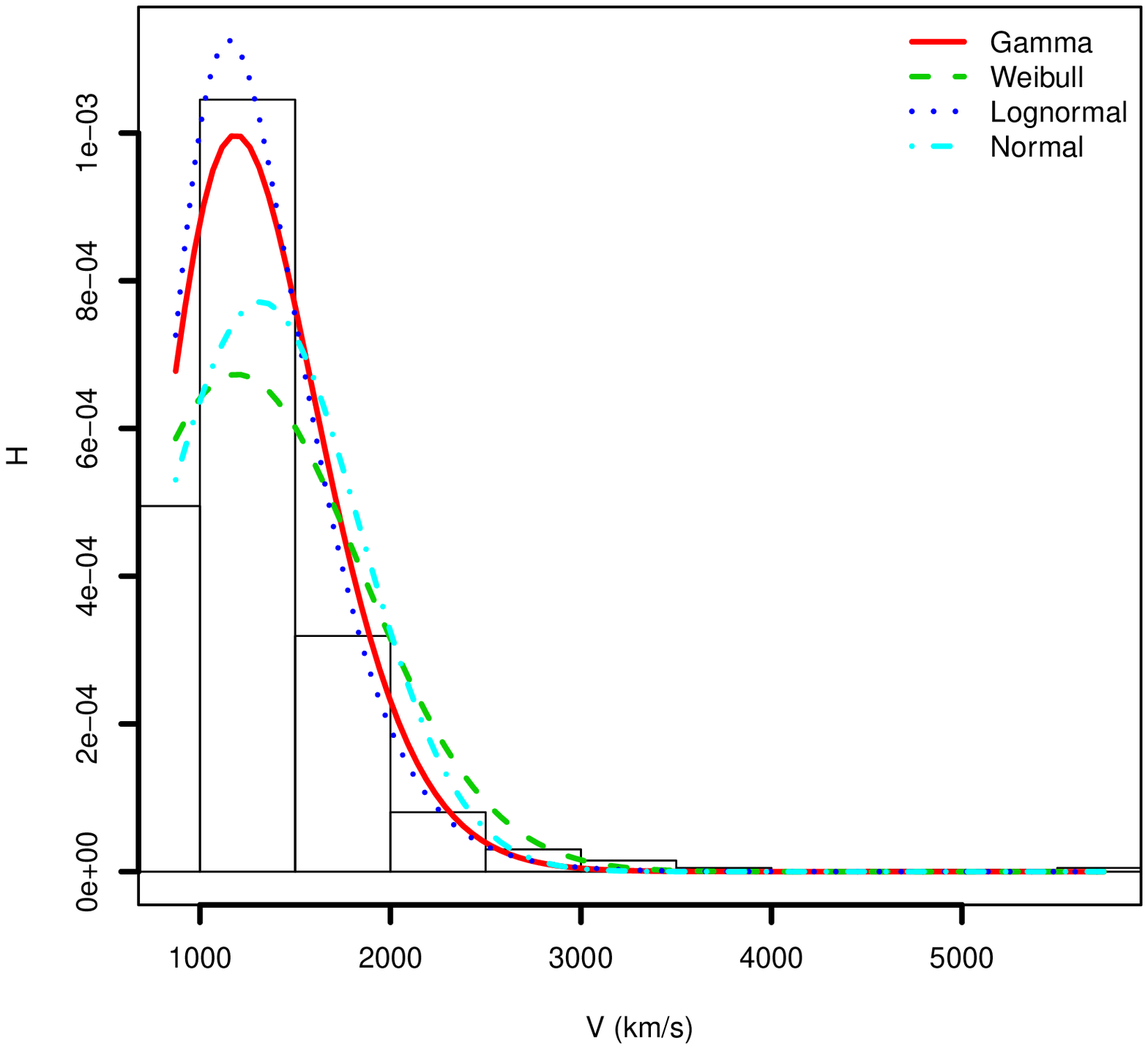}}
\subfigure{\includegraphics[width=8.75cm,height=7.5cm]{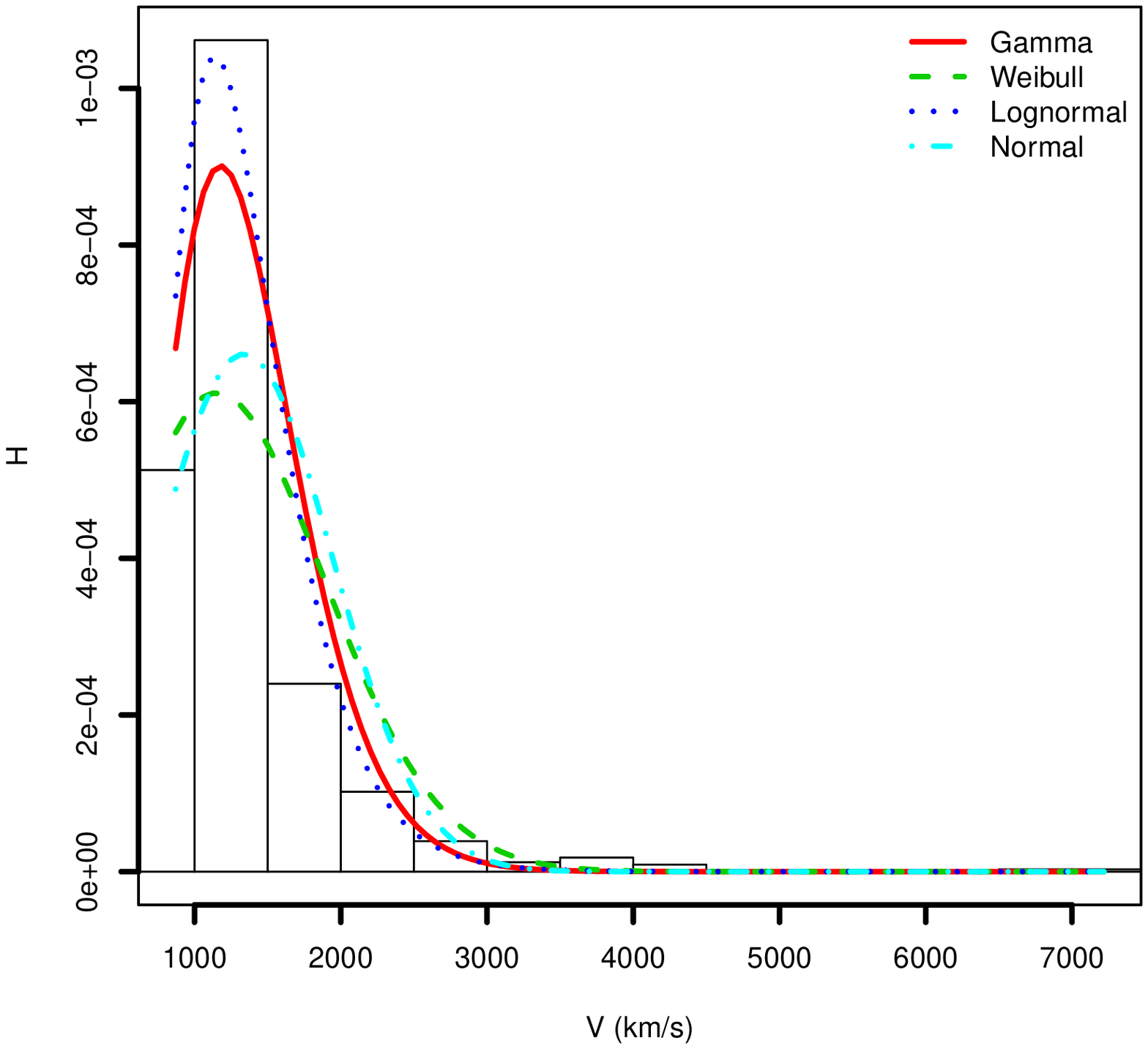}}
\end{minipage}
\caption{Distribution function of the velocity of single HVSs in models B0S1 (top-left), B025S1 (top-right), B05S1 (bottom-left), B075S1 (bottom-right). Here $H$ is normalised to area of $1$.}
\label{dvela}
\end{figure*}

\begin{figure*}
\centering
\begin{minipage}{20.5cm}
\vspace{-0.5cm}
\subfigure{\includegraphics[width=8.75cm,height=7.5cm]{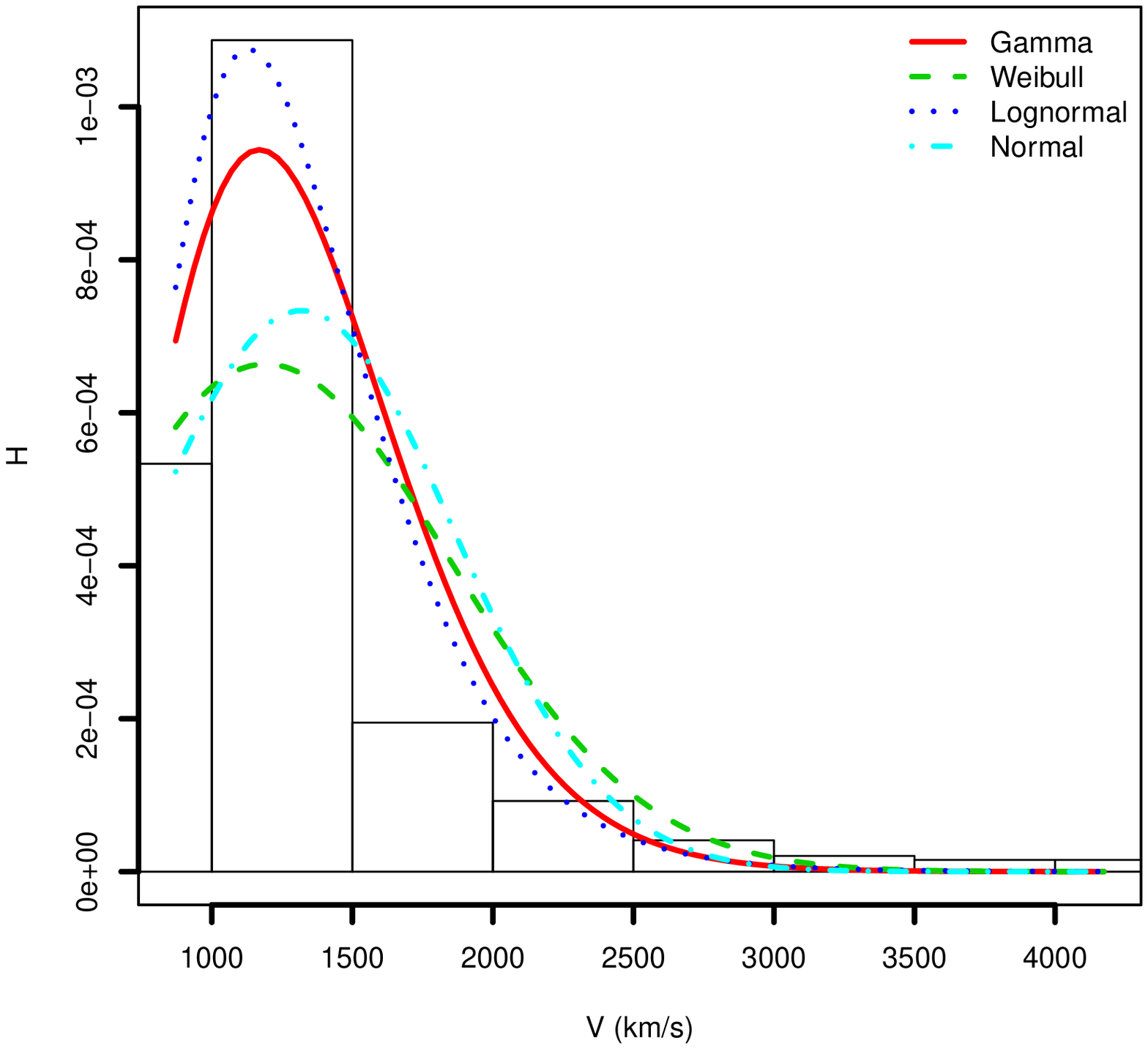}}
\subfigure{\includegraphics[width=8.75cm,height=7.5cm]{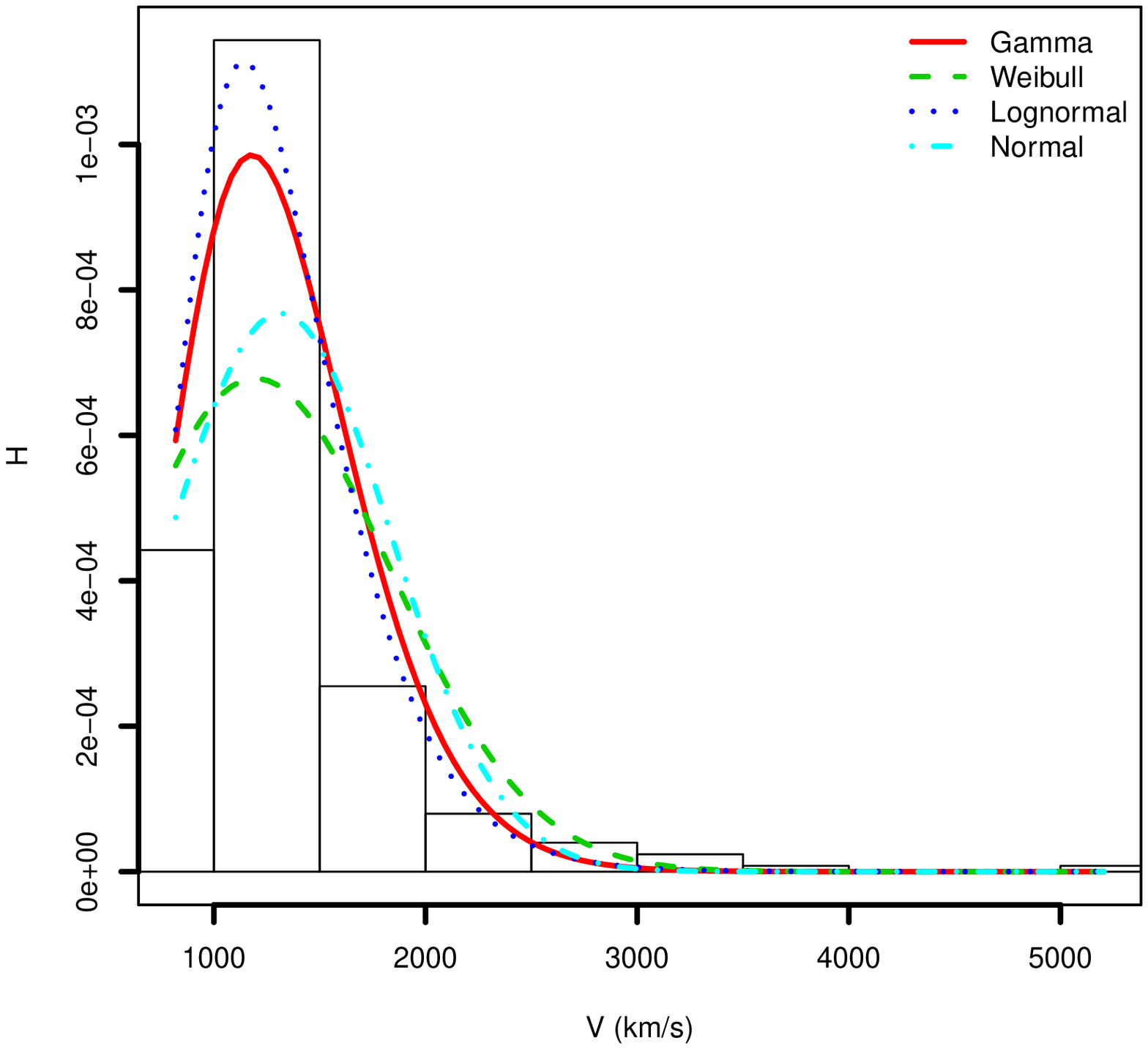}}
\end{minipage}
\caption{Distribution function of the velocity of single (left) and binary (right) HVSs for the model B1S1.}
\label{dvelb}
\end{figure*}

Figures \ref{dvela} and \ref{dvelb} show the resulting ejection velocity distributions for the models considered. Figure \ref{dvela} illustrates the resulting velocity distributions of single HVSs for different initial binary fractions of the YSC. We included in the distribution all the stars with velocity beyond the local escape speed $v_{es}(r)$, as determined by the Galactic model (see Section 2.1), and consider their velocity at the reference distance of $1$ pc from the central BH. The local (at $1$ pc) escape speed is $v_{es}\approx 940$ km s$^{-1}$, which corresponds to the cutoff in the plots. The distributions extend up to $\approx 4000-5000$ km s$^{-1}$.  We fitted our data with $4$ different functions, i.e the gamma, weibull, lognormal and normal functions. The Kolmogorov-Smirnov statistical test favours the normal distribution. Figure \ref{dvelb} shows the velocity distributions of single and binary HVSs for the model B1S1, along with the same fits of Fig. \ref{dvela}. Also in this case, the best fit is given by the normal distribution.

\begin{figure}
\centering
\includegraphics[width=8.75cm,height=7.5cm]{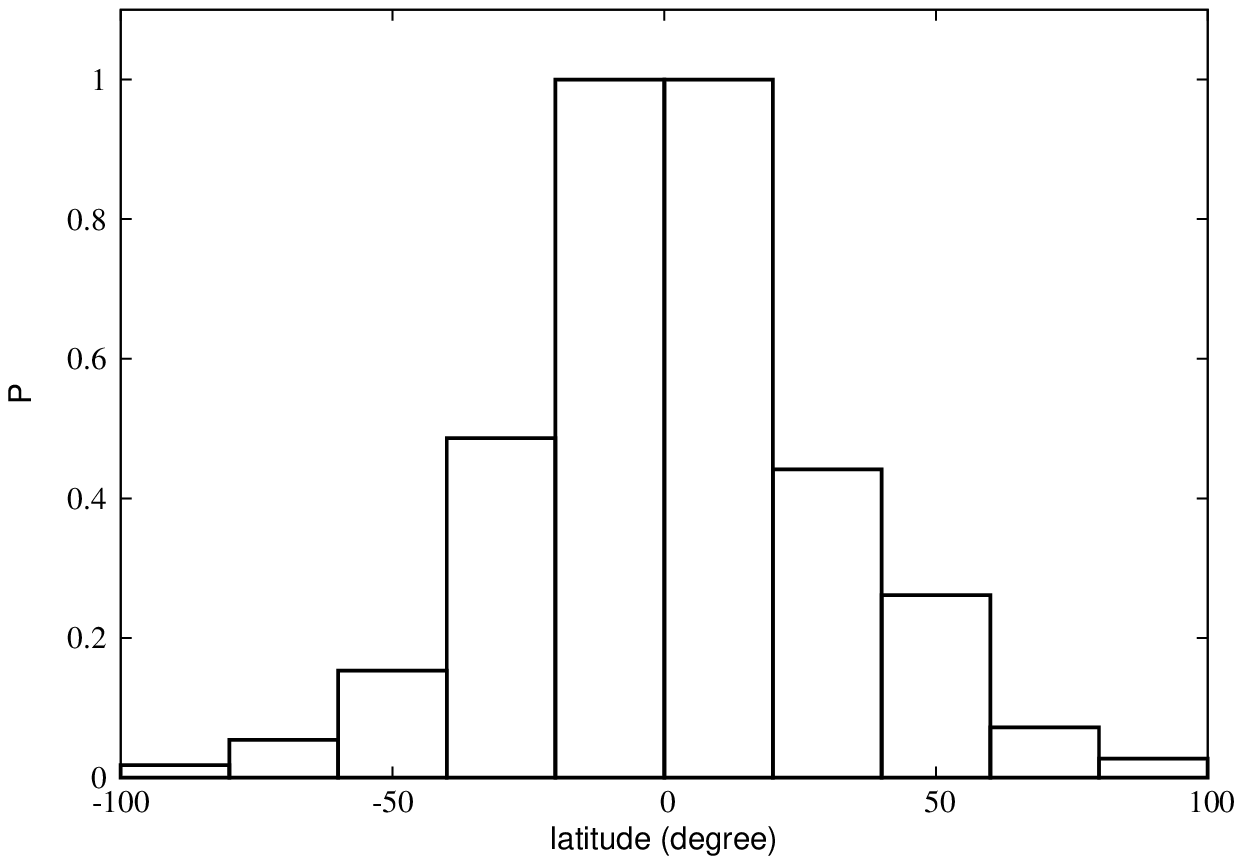}
\includegraphics[width=8.75cm,height=7.5cm]{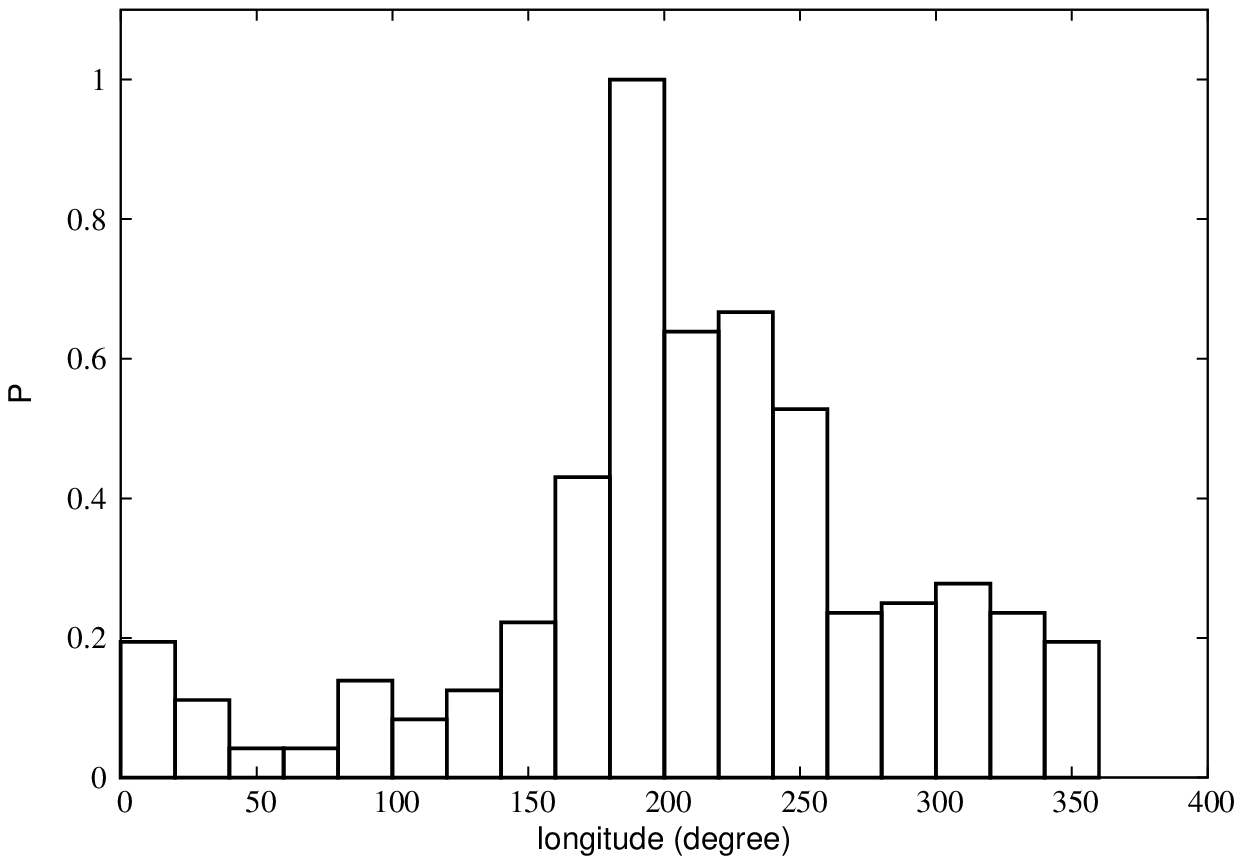}
\caption{Latitude and longitude of single HVSs for the model B1S1. Here $P$ is normalised to the maximum value.}
\label{latlong}
\end{figure}

\citet{cap15} demonstrated that another important feature of the high velocity stars originated as a consequence of the interaction of a globular cluster and a massive BH is the significant level of collimation of the ejected stars. As an illustrative example, Figure \ref{latlong} shows the distributions of the Galactocentric latitude and longitude of single HVSs for the model B1S1. As mentioned above, the cluster center-of-mass is set on the Galactic x-axis at $100$ pc off the MW center and has an initial velocity $2$ km s$^{-1}$ along the Galactic positive y-axis. As a consequence of the chosen initial conditions, the cluster center-of-mass orbital motion lies in the Galactic disk. Figure \ref{latlong} (top panel) shows that initially the latitude distribution is peaked at $0^{\circ}$. This result means that the HVSs are ejected nearly in the initial cluster orbital plane. On the other hand, the longitude distribution of HVSs quantifies the ejection direction with respect to the orbital motion of the cluster. \citet{cap15} demonstrated that high velocity stars are collimated and preferentially ejected near the cluster pericenter. Figure \ref{latlong} (bottom panel) shows the resulting longitude distribution. The distribution presents a peak at $\approx 190^{\circ}$, while the cluster center-of-mass velocity at pericenter has $\theta=180^{\circ}$. This indicates that HVSs are ejected near the pericenter along the direction of motion of the cluster.

Our results suggest that this mechanism produces single and binary HVSs in burst-like events (when the cluster is near pericenter), which will move on the orbital plane and in the direction of the cluster orbital motion. Moreover, the binary fraction of these HVS "jets" depends on the initial binary fraction in the progenitor cluster.

\begin{figure}
\centering
\includegraphics[width=8.75cm,height=7.5cm]{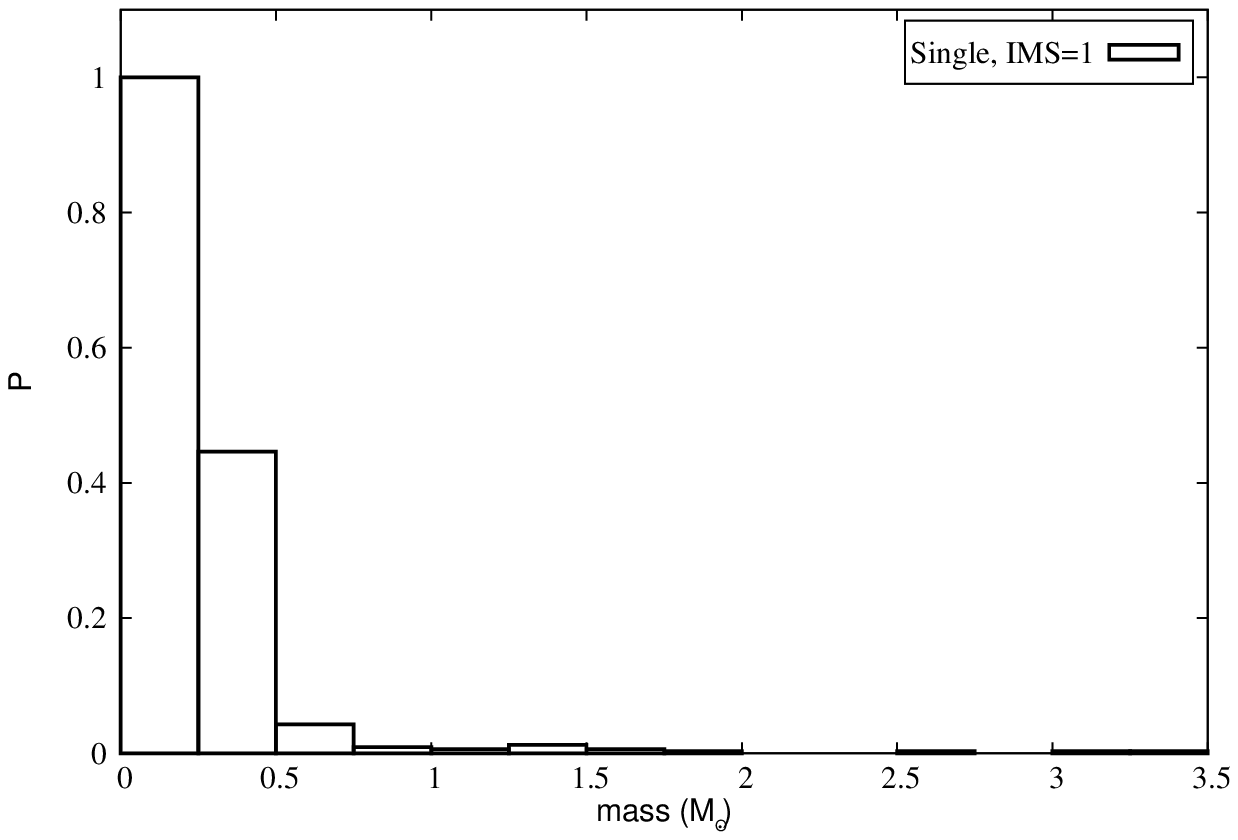}
\includegraphics[width=8.75cm,height=7.5cm]{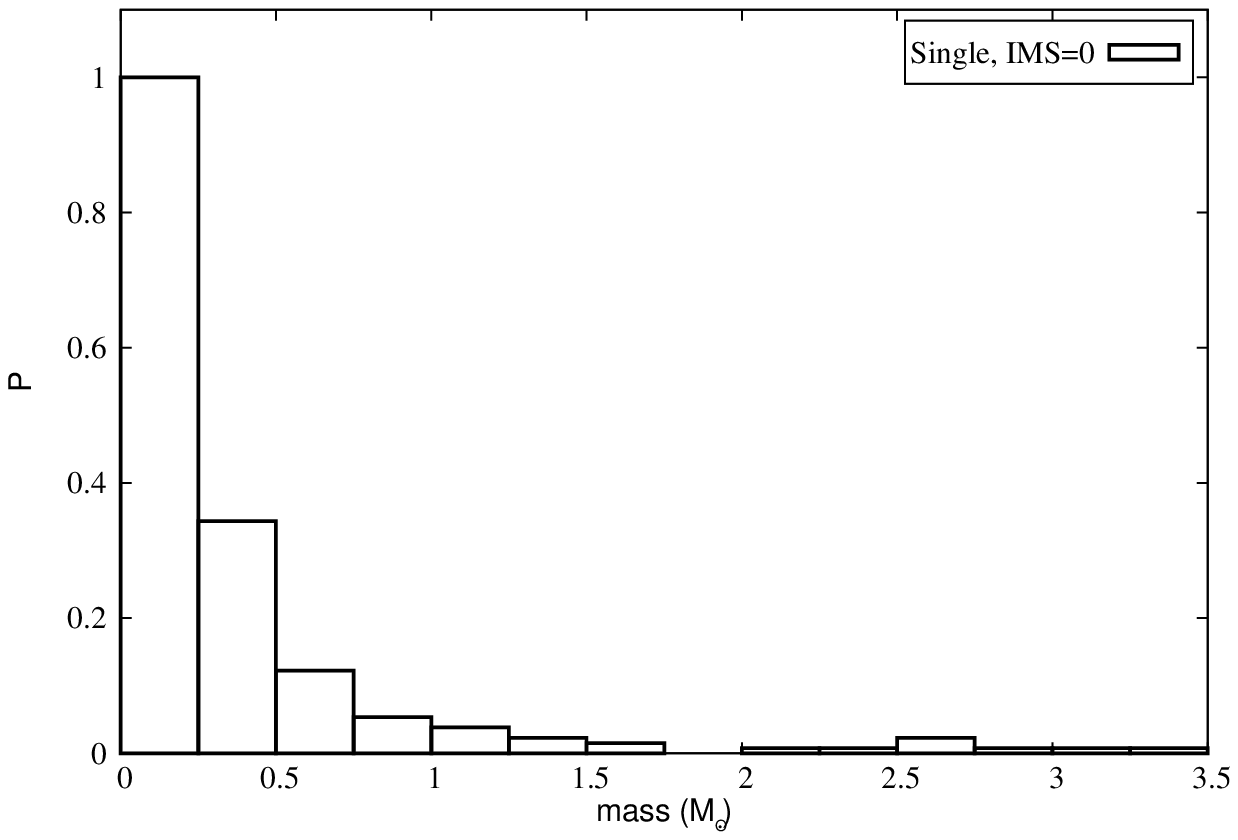}
\caption{Masses of HVSs with (top, model B1S1) and without initial mass segregation (bottom, model B1S0).}
\label{hvsmass}
\end{figure}

We investigated also the role of the initial mass segregation comparing results of model B1S1, fully mass segregated, with those of model B1S0, not mass segragated. We see that the segregation affects the mass distribution of HVSs. Figure \ref{hvsmass} shows the mass distribution of the ejected single HVSs for the cases under consideration. For the case of the fully segregated YSC, $\approx 97$\% of ejected HVSs have $m_*\lesssim 0.5$ M$_{\odot}$. Whereas, for an unsegregated YSC, the amount of stars with $m_*\gtrsim 0.5$ M$_{\odot}$ is not negligible ($\approx 20$\% of all HVSs). It is interesting to note that our model predicts the production of low-mass HVSs, as a consequence of the fact that the initial average stellar mass in the YSC is $\approx 0.5$ M$_{\odot}$. However, the observed HVSs are probably main sequence B stars with $m_*\approx 3-4$ M$_{\odot}$ \citep{brw14}. Hence, if HVSs originated from the infall of a YSC, the initial mass segregation fraction had to be $S\approx 0$. \citet{brw08} showed that HVSs are clustered in the direction of the constellation Leo (see also \citep{brw14}). \citet*{brw12} suggested that the HVSs anisotropy may reflect the anisotropy of the Milky Way gravitational potential. In the \citet{sub16} model, the anisotropy reflects the disc inclination from which HVSs are generated. In our model, the anisotropy of HVSs is a natural consequence of the cluster motion at the pericentre, while the flight times of the HVSs clumped around the constellation Leo can be described by successive bursts of HVSs produced by successive interactions between the cluster and the BH.

\subsection{The effect of a top-heavy IMF}

\citet*{dem07} showed for the first time that clusters depleted in low-mass stars have a low concentration, which is incompatible with a canonical \citet{kro01} IMF modelled by standard secular two-body relaxation evolution \citep{lei12}. \citet*{mar08} proposed a residual-gas expulsion scenario to solve this inconsistency \citep{mar10}. In this scenario, the quick gas removal from compact and primordially mass-segregated clusters leads to low-concentration clusters which are depleted in low-mass stars. Concerning stars with masses $\gtrsim 1$ M$_{\odot}$, their content depends on the ambient star-forming conditions. \citet{elm03} demonstrated that densely-packed stars would produce top-heavy IMFs in the most massive and dense clusters. It may be difficult to observe evidence for a top-heavy IMF, since stars with mass $\gtrsim 1$ M$_{\odot}$ have evolved away from the main sequence in those globular clusters, where a primordial top-heavy IMF is expected to be found \citep{mkd12}. However, some YSCs show evidence of a primordial top-heavy IMF, such as the Arches cluster (located about $25$ pc in projected distance from the GC). A top-heavy IMF can be described by
\begin{equation}
\xi(m)=
\begin{cases}
h_1\left(\frac{m}{0.08}\right)^{-1.3}& \text{$m_{min}\le m/\mathrm{M}_\odot\leq 0.50$},\\
h_2\left(\frac{m}{0.5}\right)^{-2.3}& \text{$0.50\le m/\mathrm{M}_\odot\leq 1.0$},\\
h_3\left(\frac{m}{1.0}\right)^{-\alpha_3}& \text{$1.0\le m/\mathrm{M}_\odot\leq m_{max}$},
\end{cases}
\label{eqn:imfth}
\end{equation}
where $h_1$, $h_2$ and $h_3$ are normalization factors (such that $\xi(m)$ is a continuous function) and $\alpha_3$ measures the steepness of the top-heavy IMF. Observations show that $1.65\le\alpha_3\le 2.3$ for the Arches cluster \citep{mkd12}.

\begin{figure}
\centering
\includegraphics[width=8.75cm,height=7.5cm]{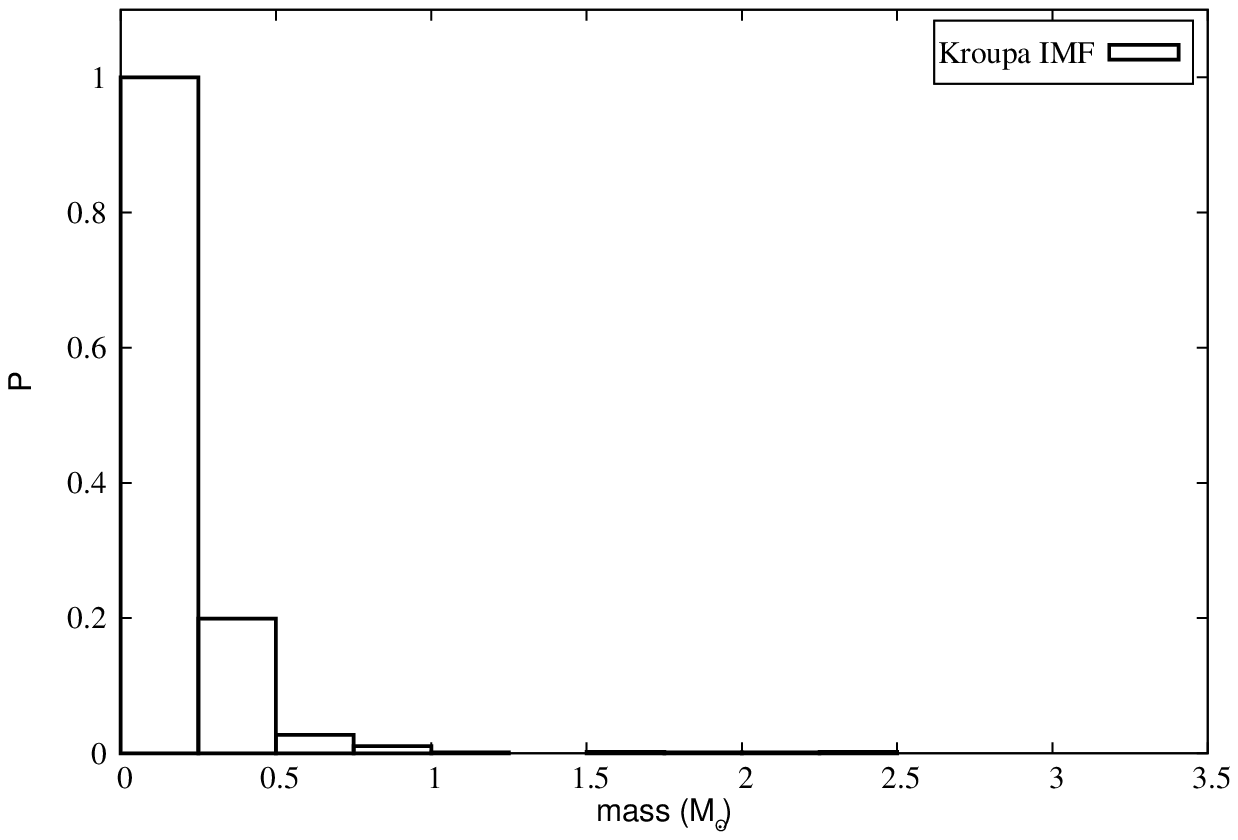}
\includegraphics[width=8.75cm,height=7.5cm]{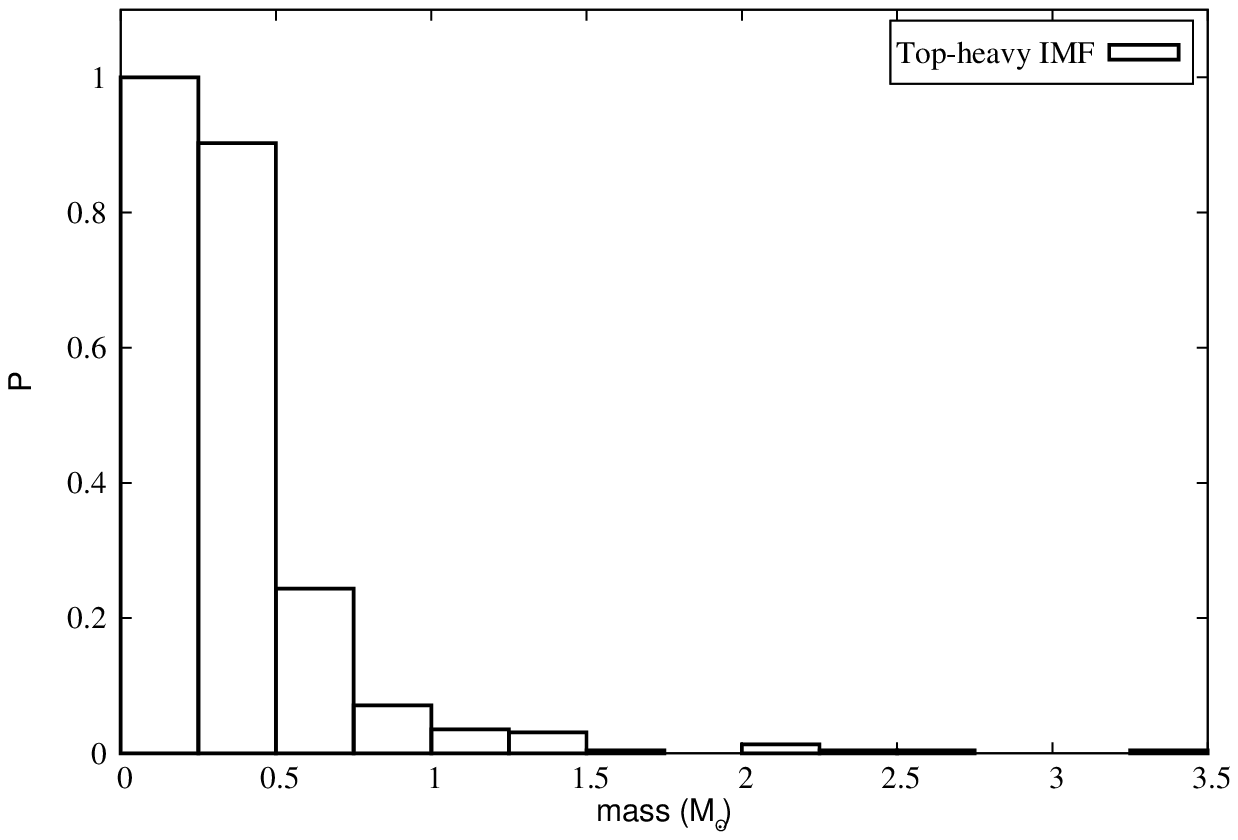}
\includegraphics[width=8.75cm,height=7.5cm]{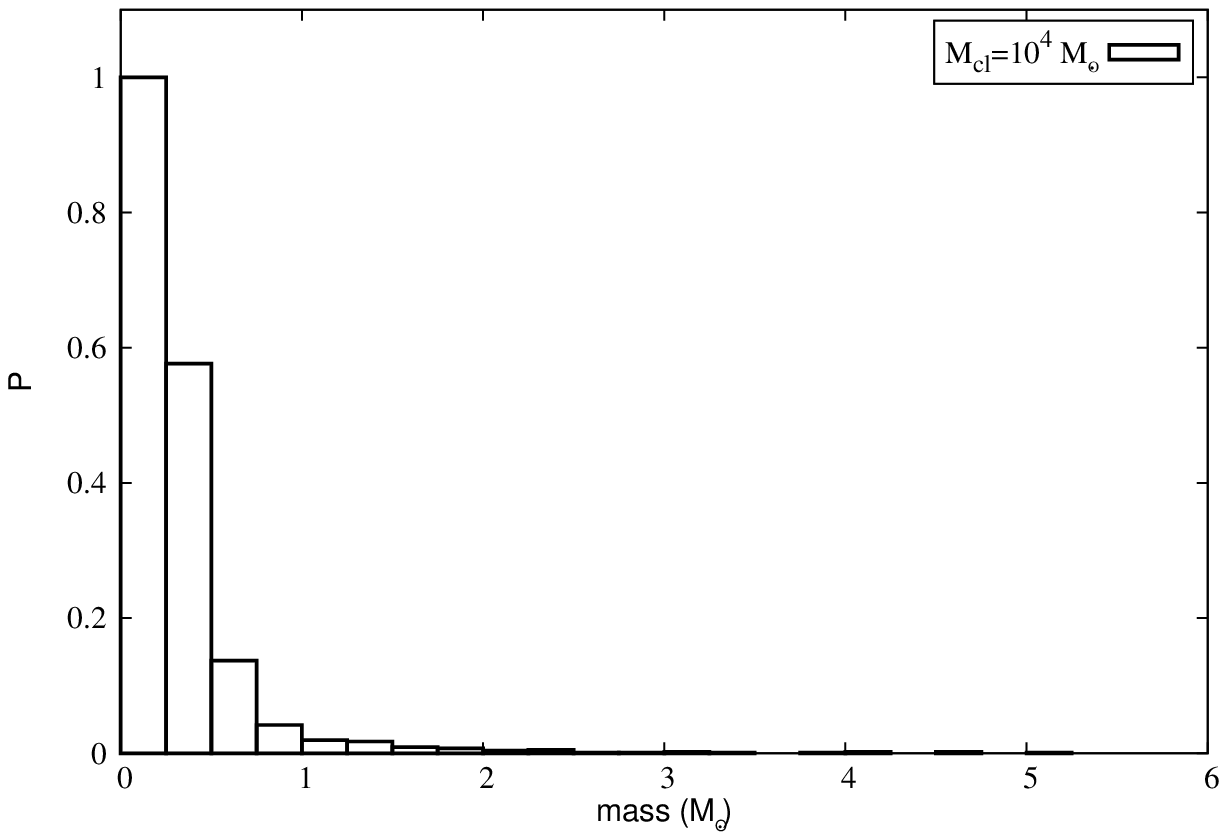}
\caption{Masses of HVSs with canonical Kroupa IMF (top, model B0S1) and top-heavy IMF (centre, model B0S1) for the case $M_{cl}=10^3$ M$_{\odot}$, and masses of HVSs with canonical Kroupa IMF (top, model B0S1) for the case $M_{cl}=10^4$ M$_{\odot}$.}
\label{topheavy}
\end{figure}

In this section we show the effects of a top-heavy IMF on the originated HVSs. We compare the results for the model B0S1 with the same model, in which the masses of the stars are sampled from a top-heavy IMF, i.e. for the same cluster mass an IMF with more massive stars than expected from the canonical \citep{kro01} IMF.

In our simulations we use Eq. \ref{eqn:imfth} with $\alpha_3=1.65$ to sample stellar masses. We find that the presence of a primordial top-heavy IMF influences the average mass of the produced HVSs. Figure \ref{topheavy} shows the mass distribution of the ejected HVSs for the cases under consideration. As discussed above, while for the case of a canonical IMF $\approx 97$\% of the ejected HVSs have $m_*\lesssim 0.5$ M$_{\odot}$ (top panel), the amount of stars with $m_*\gtrsim 0.5$ M$_{\odot}$ is not negligible ($\approx 18$\% of all HVSs) when considering a top-heavy IMF (central panel). Figure \ref{topheavy} illustrates also the mass distribution of the ejected HVSs for the case $M_{cl}=10^4$ M$_{\odot}$. In this case, $r_h=0.33\ \mathrm{pc}$ \citep{mar12} and $m_{max}\approx 110.42$ M$_\odot$ in Eq. \ref{eqn:imf} \citep{wei04,pfl07}. Figure \ref{topheavy} shows that the amount of stars with $m_*\gtrsim 0.5$ M$_{\odot}$ is not negligible ($\approx 15$\% of all HVSs) when considering a $M_{cl}=10^4$ M$_{\odot}$ cluster (bottom panel). Figure \ref{topheavy} also illustrates that the maximum mass of the ejected HVSs increases when dealing with more massive clusters, as a consequence of the \citet{wei04} relation used for $m_{max}$ (we found HVSs with masses up to $\approx 15$ M$_{\odot}$). The effect of a top-heavy IMF and of a more massive cluster is comparable to the one induced by primordial mass segregation concerning the average mass of ejected HVSs. As shown, initially not segregated clusters lead to a not negligible fraction of massive HVSs. If some of the observed HVSs originated from the infall of a YSC, three possible explanations are feasible, i.e. the cluster initial mass segregation $S\approx 0$, or the cluster primordial IMF was top-heavy, or $M_{cl}\gtrsim 10^4$ M$_{\odot}$. Moreover, the efficiency in converting cluster stars into HVSs is about one order of magnitude larger with respect to the $10^3$ M$_{\odot}$ cluster (see also \citet{cap15}).

\begin{figure}
\centering
\vspace{-0.5cm}
\includegraphics[width=8.75cm,height=7.5cm]{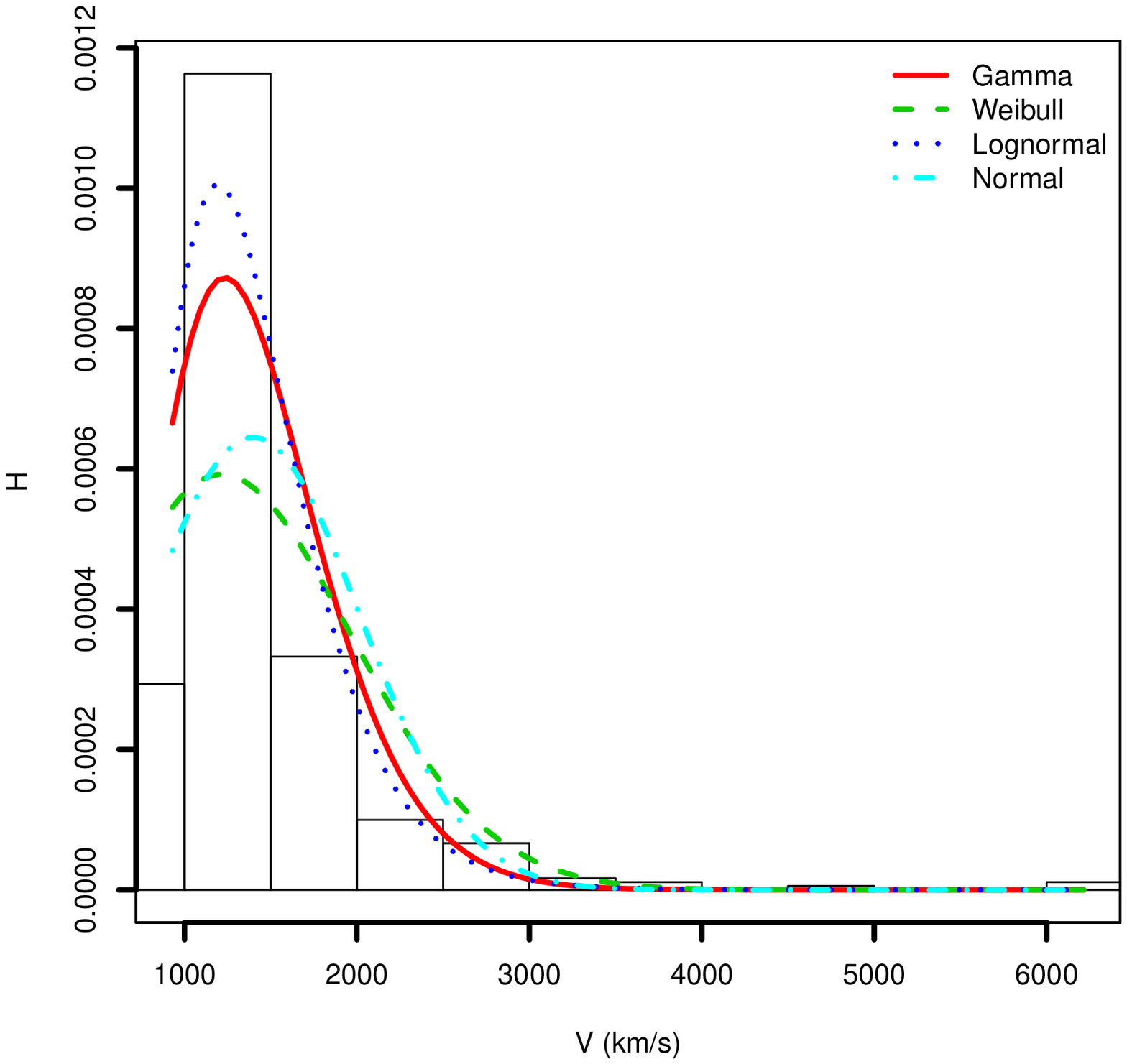}
\includegraphics[width=8.75cm,height=7.5cm]{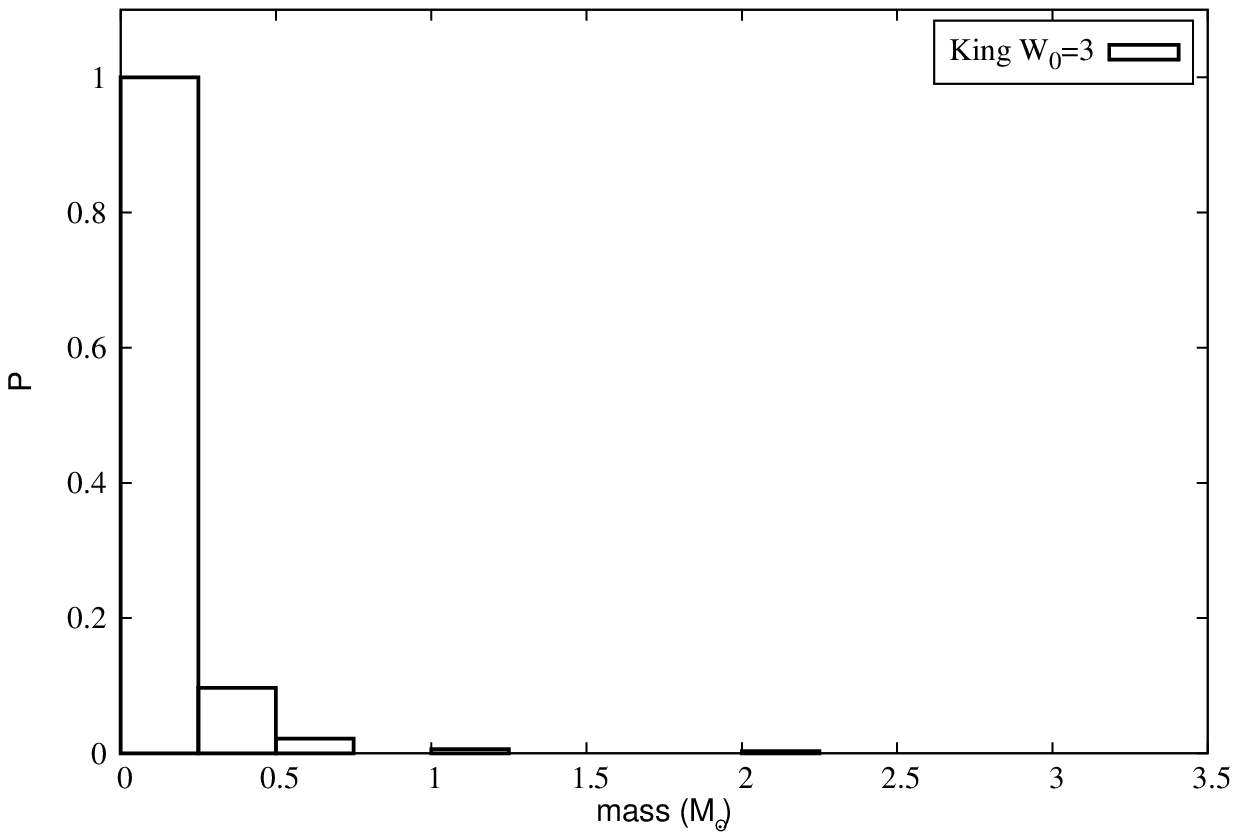}
\caption{Distribution function of the velocity of HVSs (top) and their masses (bottom) for the model B0S1 for the case the cluster density profile is modelled with a \citet{kin66} profile with $W_0=3$ and $r_h=0.25$ pc. There is no statistically significant difference to the Plummer case (Fig. \ref {dvela} (top-left panel) and Fig. \ref{topheavy} (top panel).)}
\label{kingw3}
\end{figure}

Concerning the Arches cluster, \citet*{har10} showed that, upon comparison of simulations with observations, it may be described by a \citet{kin66} model with a value of the central dimensionless potential $W_0=3$. We perform simulations for the model B0S1, whose star distribution is sampled from a \citet{kin66} profile with $W_0=3$. Figure \ref{kingw3} shows the distribution function of the velocity of the produced HVSs and their masses. We find that there are no significant differences from the case where stars are sampled from a \citet{plu11} profile. We note here that the Plummer density distribution function is the simplest solution to the collisionless Boltzmann equation, that it well describes the simplest stellar dynamical systems \citep{plu11}, and that it evolves to a King profile due to the energy equipartition process and the influence of the tidal field.

\subsection{The fate of hypervelocity binary stars}

We study the fate of the ejected binary HVSs. As discussed, binary HVSs can undergo three different fates. The binary star can survive travelling farther at hyper velocity, can merge or can be disrupted as a consequence of the velocity kick.

\begin{figure}
\centering
\includegraphics[width=8.75cm,height=7.5cm]{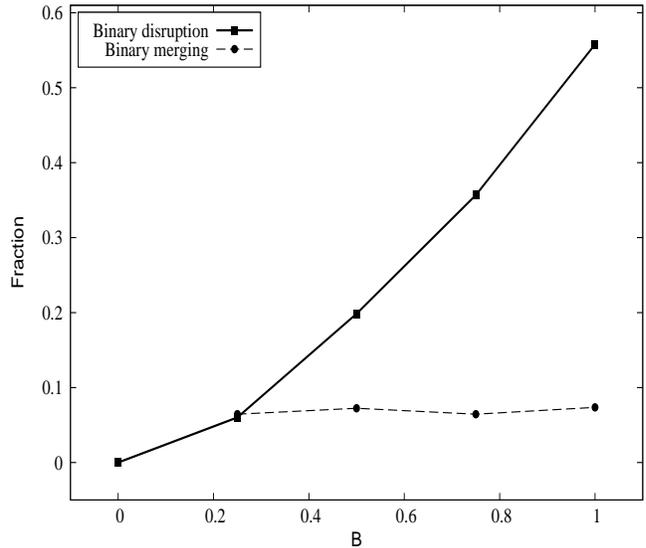}
\caption{Disrupted and merged binary HVSs.}
\label{dismer}
\end{figure}

Figure \ref{dismer} shows the fraction of single HVSs that originates as consequence of the binary disruption discussed in the previous section. This fraction increases when the initial binary fraction is larger. For the case $B=1$ about $60 \%$ of single HVSs are generated through this channel.

\begin{figure}
\centering
\includegraphics[width=8.75cm,height=7.5cm]{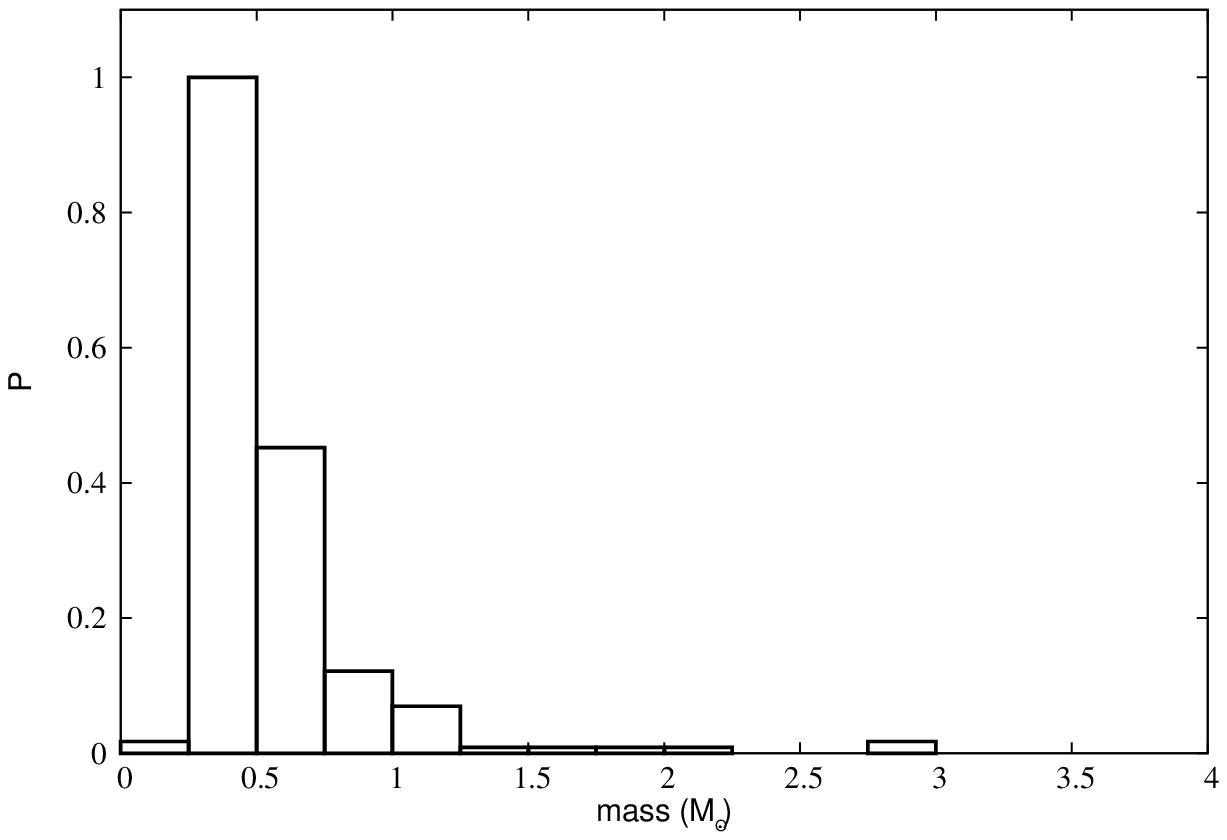}
\includegraphics[width=8.75cm,height=7.5cm]{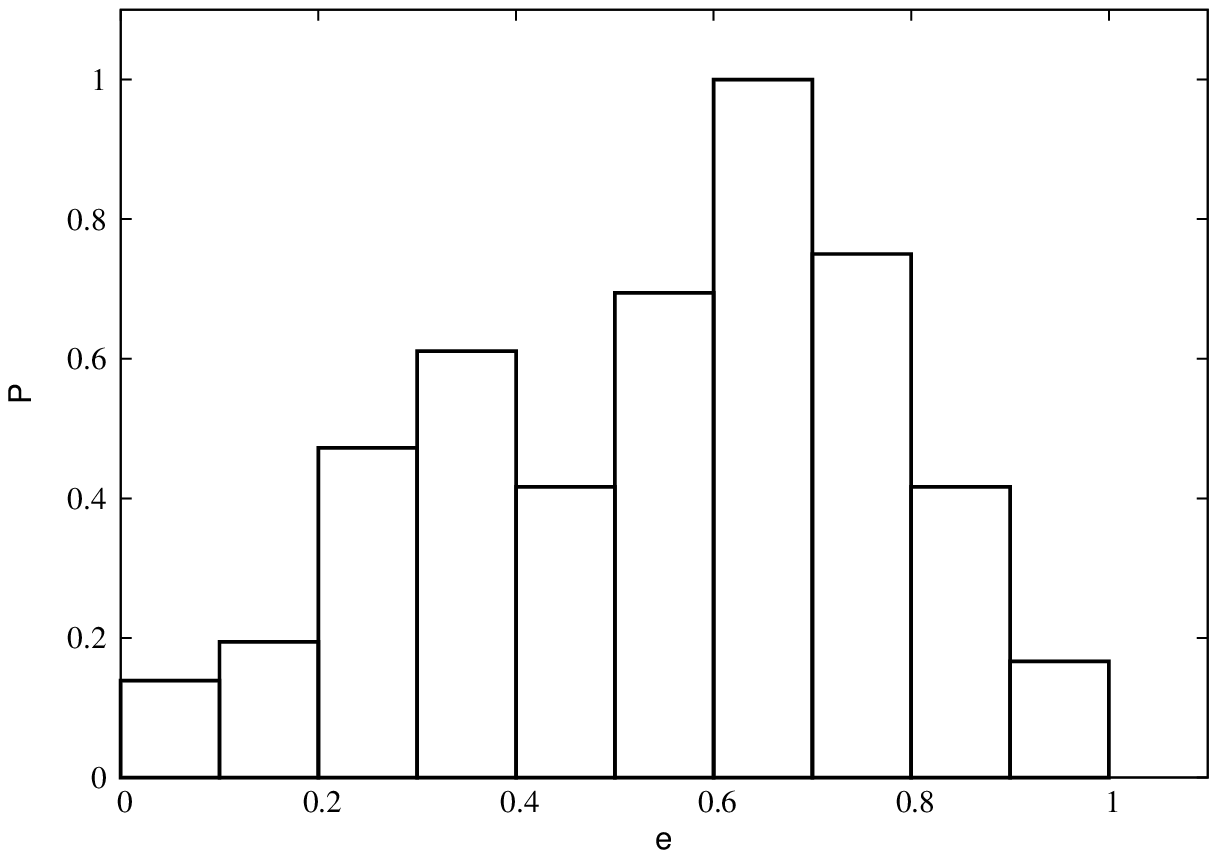}
\includegraphics[width=8.75cm,height=7.5cm]{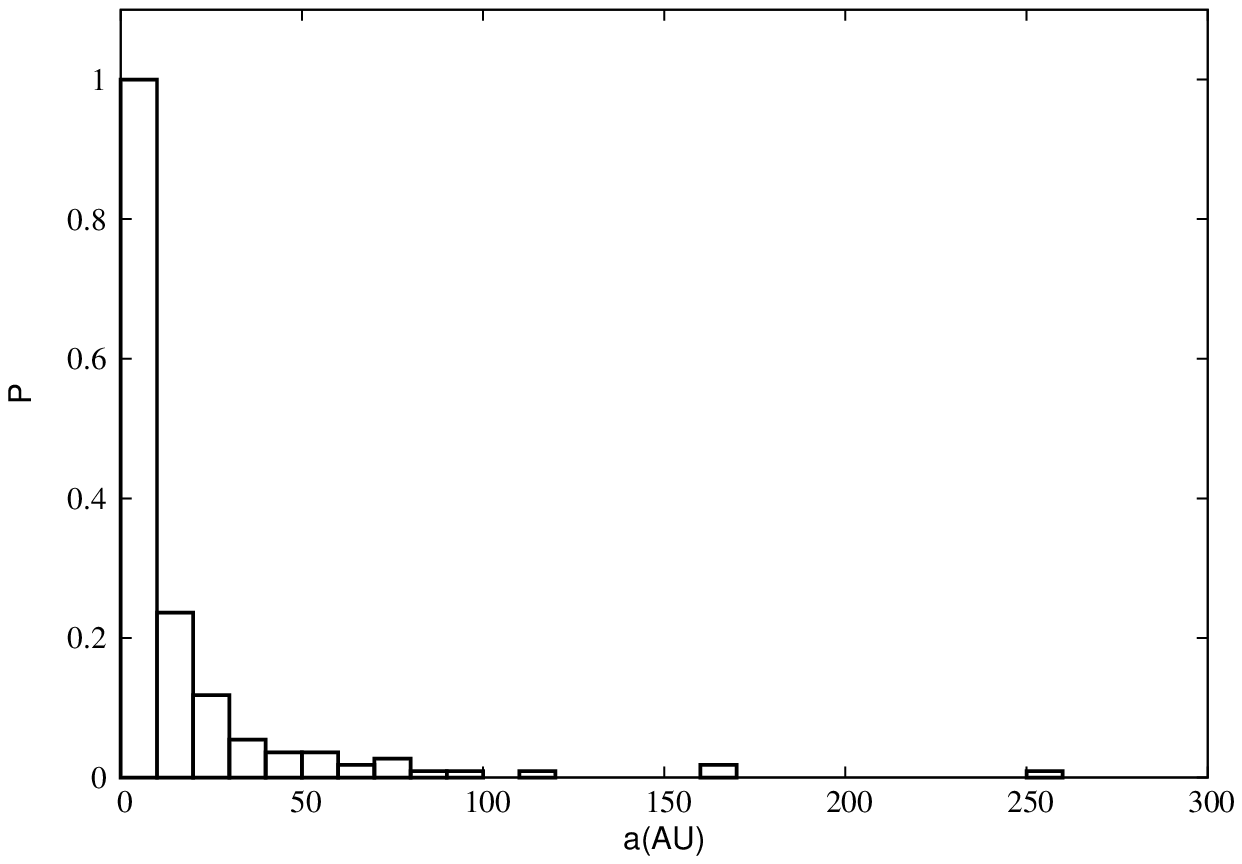}
\caption{Mass (top), eccentricity (centre) and semi-major axis (bottom) distributions of binary HVSs for the model B1S1.}
\label{meccsem}
\end{figure}

Figure \ref{meccsem} shows the mass, eccentricity and semi-major axis of binary HVSs for the model B1S1. Due to the initial choice of mass sampling from the canonical IMF \citep{kro01,kro13}, the mass distribution (top panel) is peaked at $m_*\lesssim 0.5$ M$_{\odot}$. The eccentricity distribution (central panel) shows that ejected binaries are highly eccentric, with a peak at $e\approx 0.7$. The semi-major axis distribution (bottom panel) indicates that most of the ejected binary HVSs have $a\lesssim 100$ AU. Because the average stellar mass in the YSC is $\approx 0.5$, the total mass of the binary $\approx 1$ M$_{\odot}$, and $P\propto a^{3/2})$, the period of binary HVSs results to be $P\lesssim 10^3$ days. Figure \ref{meccsem} thus indicates, so, that the binary HVSs ejected from an infalling YSC are generally compact and eccentric.

Such eccentric and compact binary stars can merge while travelling across the Galaxy at hypervelocities. If we assume that the evolution of a binary is described by the \citet{kro95b} eigenevolution, the evolution of orbital elements occurs when the stars of the binary are at the pericentre 
\begin{equation}
R_{per}=(1-e)\left(\frac{P_b}{1\ \mathrm{day}}\right)^{2/3}\left(\frac{m_1+m_2}{1\ \mathrm{M}_{\odot}}\right)^{1/3}\ \mathrm{AU},
\end{equation}
where $e$ is the binary eccentricity, $P_b$ is the binary period in days and $m_1$ and $m_2$ are the masses of the primary and secondary, respectively. According to \citet{kro95b}, the binary eccentricity evolves. after the ejection from the cluster, according to
\begin{equation}
\frac{\dot{e}}{e}=-\dot{\rho},
\end{equation}
where
\begin{equation}
\rho=\left(\frac{\lambda R_{\odot}}{R_{per}}\right)^{\chi}.
\end{equation}
In the previous equation, $R_{\odot}=4.6523\times 10^{-3}$ AU is the Sun's radius, while $\lambda$ indicates the length scale over which significant evolution of the orbital elements occurs, $\chi$ measures the interaction strength of the two stars \citep{kro95b}. If eigenevolution occurs during pre-main sequence $\lambda=28$ and $\chi=0.75$, whereas $\lambda_{ms}=24.7$ and $\chi_{ms}=8$ for the case of main sequence phase. The initial binary period $P_{b,in}$ evolves to
\begin{equation}
P_{b,fin}=P_{b,in}\left(\frac{m_{tot,in}}{m_{tot,fin}}\right)^{1/2}\left(\frac{1-e_{in}}{1-e_{fin}}\right)^{3/2},
\end{equation}
where $m_{tot,in}$ and $m_{tot,fin}$ are the initial and final total mass, respectively. On the other hand, the initial mass ratio $q_{in}=m_1/m_2$ is assumed to evolve to
\begin{equation}
q_{fin}=q_{in}+(1-q_{in})\rho^{*}
\end{equation}
where
\begin{equation}
\rho^{*}=
\begin{cases}
\rho& \text{$\rho\le 1$}\\
1& \text{$\rho> 1$}
\end{cases}
\end{equation}
Finally, the final mass of the secondary will be $m_{2,fin}=q_{fin}m_{1,in}$, while $m_{1,fin}=m_{1,in}$. The components of the binary are considered merged if their semi-major axis after eigenevolution is $\lesssim 10\ R_{\odot}$ \citep{kro95b}.

We apply the eigenevolution to the ejected binary HVSs to quantify how many of them will merge. Figure \ref{dismer} shows the fraction of binaries that merge, as a consequence of the eigenevolution, as a function of the initial binary fraction. The same binaries merge both if we use pre-main sequence $\lambda$,$\chi$ and main sequence $\lambda_{ms}$,$\chi_{ms}$ since merging binaries are very compact ($a\lesssim 0.2$ AU) and eccentric ($e\gtrsim 0.7$). We find that the fraction of merged binaries is nearly constant ($\approx 7 \%$) independent of the initial $B$. This small fraction of HVSs will continue their paths in the Galactic field as blue stragglers.

\section{Conclusions}
\label{sec:con}

In this paper we investigated the origin of HVSs as a consequence of the close passage of a YSC around the Milky Way's central SMBH. During the close encounter, some stars are stripped from the cluster and may be ejected with high velocities \citep{cap15,fra16}. We focussed our attention on the stars lost by the system and ejected at hyper velocities after the interaction with the BH, examining the role of the cluster initial binary fraction and mass segregation.

We found that this mechanism produces single and binary HVSs in a burst-like event (when the cluster is at the orbital pericenter), nearly in the initial cluster orbital plane and in the direction of the cluster orbital motion. The binary fraction of these HVSs jets depends on the initial fraction of binaries in the progenitor cluster. On the other side, the initial mass segregation affects the mass of ejected stars: the smaller the initial segregation fraction, the larger the average mass of HVSs. Moreover, we found that also the top-heaviness of the IMF and the total mass of the cluster affects the mass distribution of HVSs: a top-heavy IMF and a large cluster mass increase the average mass of HVSs.

We also quantified how many binary stars survive at hypervelocities. Applying the \citet{kro95b} eigenevolution formulas we found that $\approx 7$\% of binary HVSs will merge to become, eventually, blue straggler HVSs. Both binary and blue stragglers HVSs have been observed. \citet{ede05} observed a star moving with a velocity of at least $530$ km s$^{-1}$ in the Galactic rest frame. According to stellar atmosphere fits the star is a 9 M$_{\odot}$ main-sequence star $50$ kpc away. The lifetime of this star is several times shorter than its flight time from the Milky Way, suggesting an LMC origin \citep{gua07} or a blue straggler origin \citep{per09}. The latter channel suggest that the progenitor was likely a binary system ejected from the Milky Way at $\gtrsim 800$ km s$^{-1}$ \citep{brw15}. Recently, \citet{nem16} have spotted the first binary HVS candidate $\approx 5.7$ kpc far from the GC travelling at $\approx 571$ km s$^{-1}$.

\citet{brw08} showed that HVSs are clustered in the direction of the constellation Leo. \citet*{brw12} proposed that the HVS anisotropy could reflect the anisotropy of the Galactic potential. Our model predicts that the anisotropy of HVSs is a natural consequence of cluster motion at the pericentre. This is natural if most stars form in clusters \citep{kro05}.

\citet*{paw12} showed that the MW is surrounded by a disk of mostly coorbiting satellite galaxies, the vast polar structure (VPOS) \citep*{ktb05}. Is the disk of stars around the super massive BH of the MW, and/or some gaseous disk there, oriented as the VPOS? The known HVSs do correlate with the VPOS \citep*{paw13} and if the innermost accretion disk or circum-nuclear disk of the super massive BH is aligned with the VPOS, then results showed in this paper may nicely explain the anisotropy of the flux of HVSs: a moderate embedded cluster forms in the gaseous disk and falls towards the super massive BH and produces the HVS flux. This is similar to the dwarf-galaxy picture \citep{aba09}, but is more plausible since stars form in embedded clusters, also in the inner Galaxy, and the youth of the HVSs is then also less of a problem \citep{brw14}.

HVSs are powerful tools to investigate the physics of the GC, being in an accessible region of the sky, and also to study the dark sector of our Galaxy \citep{per09,frl16}. When a star cluster passes near the SMBH, it produces HVSs as a consequence of the strong interaction with the SMBH tidal field. If HVSs are generated through the process presented in this paper, by studying their spatial and velocity distribution, it is possible to constrain the physical properties of clusters that have infallen onto the central SMBH, over the course of the Milky Way's history, including their binary and mass composition.

\section*{Acknowledgments}

GF acknowledges hospitality from Sambaran Banerjee and Pavel Kroupa, and the University of Bonn, where the early plan for this work was conceived. GF acknowledges Sverre Aarseth for fruitful discussions on the use of the code \textsc{nbody6}. GF acknowledges Andreas K\"{u}pper and Yohai Meiron for useful and helpful discussions and comments about the contents investigated in this paper. Part of this work was performed at the Aspen Center for Physics, which is supported by National Science Foundation grant PHY-1066293. At this regard, RCD thanks the Simons foundation for the grant which allowed him a period of stay at the Aspen Center for Physics where he developed part of this work.

\label{lastpage}


\begin{thebibliography}{99}
\bibitem[\protect\citeauthoryear{Aarseth}{2003}]{aar03} Aarseth S.J., 2003 Gravitational N-body simulations: tools and algorithms. Cambridge University Press
\bibitem[\protect\citeauthoryear{Abadi, Navarro \& Steinmetz}{Abadi et al.}{2009}]{aba09} Abadi M.G., Navarro J.F., Steinmetz M., 2009, ApJ Lett., 691, L63
\bibitem[\protect\citeauthoryear{Arca-Sedda, Capuzzo-Dolcetta \& Spera}{Arca-Sedda et al.}{2016}]{arc16} Arca-Sedda M., Capuzzo-Dolcetta R., Spera M., 2016, MNRAS, 456, 2457
\bibitem[\protect\citeauthoryear{Baumgardt, De Marchi \& Kroupa}{Baumgardt et al.}{2008}]{bau08} Baumgardt H., De Marchi G., Kroupa P., 2008, ApJ, 685, 247
\bibitem[\protect\citeauthoryear{Brown}{2015}]{brw15} Brown W.R., 2015, Annu. Rev. Astron. Astrophys., 53, 157
\bibitem[\protect\citeauthoryear{Brown, Geller \& Kenyon}{Brown et al.}{2012}]{brw12} Brown W.R., Geller M.J., Kenyon S.J., 2012b, ApJ, 751, 55
\bibitem[\protect\citeauthoryear{Brown, Geller \& Kenyon}{Brown et al.}{2014}]{brw14} Brown W.R., Geller M.J., Kenyon S.J., 2014, ApJ, 787, 89
\bibitem[\protect\citeauthoryear{Brown, Geller \& Kenyon}{Brown et al.}{2008}]{brw08} Brown W.R., Geller M.J., Kenyon S.J., Bromley B.C., 2008, ApJ Lett., 690, L69
\bibitem[\protect\citeauthoryear{Brown et al.}{2010}]{brw10} Brown W.R., Geller M.J., Kenyon S.J., Diaferio A., 2010, AJ, 139, 59
\bibitem[\protect\citeauthoryear{Brown et al.}{2005}]{brw05} Brown W.R., Geller M.J., Kenyon S.J., Kurtz M.J., 2005, ApJ Lett., 622, L33
\bibitem[\protect\citeauthoryear{Capuzzo-Dolcetta \& Fragione}{2015}]{cap15} Capuzzo-Dolcetta R., Fragione G., 2015, MNRAS, 454, 2677
\bibitem[\protect\citeauthoryear{De Marchi, Paresce \& Pulone}{De Marchi et al.}{2007}]{dem07} De Marchi G., Paresce F., Pulone L., 2007, ApJ, 656, L65
\bibitem[\protect\citeauthoryear{Edelmann et al.}{2005}]{ede05} Edelmann H., et al., 2005, ApJ Lett., 634, L181
\bibitem[\protect\citeauthoryear{Elmegreen \& Shadmehri}{2003}]{elm03} Elmegreen B.G., Shadmehri M., 2003, MNRAS, 338, 817
\bibitem[\protect\citeauthoryear{Fragione \& Capuzzo-Dolcetta}{2016}]{fra16} Fragione G., Capuzzo-Dolcetta R., 2016, MNRAS, 458, 2596
\bibitem[\protect\citeauthoryear{Fragione \& Ginsburg}{2016}]{frg16} Fragione G., Ginsburg I., 2016, preprint (arXiv:1609.03905)
\bibitem[\protect\citeauthoryear{Fragione \& Loeb}{2016}]{frl16} Fragione G., Loeb A., 2016, preprint (arXiv:1608.01517)
\bibitem[\protect\citeauthoryear{Ginsburg \& Loeb}{2006}]{gil06} Ginsburg I., Loeb A., 2006, MNRAS, 368, 221
\bibitem[\protect\citeauthoryear{Ginsburg \& Loeb}{2007}]{gil07} Ginsburg I., Loeb A., 2007, MNRAS, 376, 492
\bibitem[\protect\citeauthoryear{Ginsburg, Loeb \& Wegner}{Ginsburg et al.}{2012}]{gin12} Ginsburg I., Loeb A., Wegner G.A., 2012, MNRAS ,423.1, 948
\bibitem[\protect\citeauthoryear{Gnedin et al.}{2005}]{gne05} Gnedin O.Y., Gould A., Miralda-Escud\'{e} J., Zentner A.R., 2005, ApJ, 634, 344
\bibitem[\protect\citeauthoryear{Gnedin et al.}{2010}]{gne10} Gnedin O.Y., Brown W.R., Geller M.J., Kenyon S.J., 2010, ApJ Lett., 720, L108
\bibitem[\protect\citeauthoryear{Goodwin \& Kroupa}{2005}]{goo05} Goodwin S.P., Kroupa, P., 2005, A\& A, 439, 565
\bibitem[\protect\citeauthoryear{Gould \& Quillen}{2003}]{gou03} Gould A., Quillen A.C., 2003, ApJ, 592, 935
\bibitem[\protect\citeauthoryear{Gualandris \& Portegies Zwart}{2007}]{gua07} Gualandris A., Portegies Zwart S., 2007, MNRAS, 376, L29
\bibitem[\protect\citeauthoryear{Haas \& \v{S}ubr}{2016}]{haa16} Haas J., \v{S}ubr L., 2016, ApJ, 822, 25
\bibitem[\protect\citeauthoryear{Harfst, Portegies Zwart \& Stolte}{Harfst}{2010}]{har10} Harfst S., Portegies Zwart S., Stolte A., 2010, MNRAS, 409, 628
\bibitem[\protect\citeauthoryear{Hernquist}{1990}]{her90} Hernquist L., 1990, ApJ, 356, 359
\bibitem[\protect\citeauthoryear{Hills}{1988}]{hil88} Hills J.G., 1988, Nature, 331, 687
\bibitem[\protect\citeauthoryear{Hosek et al.}{2015}]{hos15} Hosek Jr M.W., Lu J.R., Anderson J., Ghez A.M., Morris M.R., Clarkson W.I., 2015, ApJ, 813, 27
\bibitem[\protect\citeauthoryear{Kenyon et al.}{2008}]{ken08} Kenyon S.J., Bromley B.C., Geller M.J., Brown W.R., 2008, ApJ, 680, 312
\bibitem[\protect\citeauthoryear{Kenyon et al.}{2014}]{ken14} Kenyon S.J., Bromley B.C., Brown W.R., Geller M.J., 2014, ApJ, 793, 122
\bibitem[\protect\citeauthoryear{King}{1966}]{kin66} King I. R., 1966, AJ, 71, 64
\bibitem[\protect\citeauthoryear{Kroupa}{1995a}]{kro95a} Kroupa P., 1995a, MNRAS, 277, 1491
\bibitem[\protect\citeauthoryear{Kroupa}{1995b}]{kro95b} Kroupa P., 1995b, MNRAS, 277, 1507
\bibitem[\protect\citeauthoryear{Kroupa}{2001}]{kro01} Kroupa P., 2001, MNRAS, 322, 231
\bibitem[\protect\citeauthoryear{Kroupa}{2005}]{kro05} Kroupa P., 2005, in ESA Special Publication, Vol. 576, The Three-Dimensional Universe with Gaia ed. Turon C., O’Flaherty K.S., Perryman M.A.C., Noordwijk ESA, 629
\bibitem[\protect\citeauthoryear{Kroupa}{2008}]{kro08} Kroupa P., 2008, in The Cambridge N-Body Lectures, ed. Aarseth S.J., Tout C.A., Mardling R.A. (Lecture Notes in Physics Vol. 760; Berlin: Springer), 181
\bibitem[\protect\citeauthoryear{Kroupa, Theis \& Boily}{Kroupa et al.}{2005}]{ktb05} Kroupa P., Theis C., Boily C.M., 2005, A\& A, 431, 517
\bibitem[\protect\citeauthoryear{Kroupa et al.}{2013}]{kro13} Kroupa P., Weidner C., Pflamm-Altenburg J. et al., 2013, The stellar and sub-stellar initial mass function of simple and composite populations, in Planets, Stars and Stellar Systems, Springer Netherlands, 115
\bibitem[\protect\citeauthoryear{K\"{u}pper et al.}{2011}]{kup11} K\"{u}pper A.H.W., Maschberger T., Kroupa P., Baumgardt H., 2011, MNRAS, 417, 2300
\bibitem[\protect\citeauthoryear{Leigh et al.}{2012}]{lei12} Leigh N., Umbreit S., Sills A., Knigge C., de Marchi G., Glebbeek E., Sarajedini A., 2012, MNRAS, 422, 1592
\bibitem[\protect\citeauthoryear{Leigh et al.}{2015}]{lei15} Leigh N.W.C. et al., 2015, MNRAS, 446, 226
\bibitem[\protect\citeauthoryear{Li et al.}{2015}]{li15} Li Y. et al., 2015, 15.8, 1364
\bibitem[\protect\citeauthoryear{L\"{o}ckmann, Baumgardt \& Kroupa}{L\"{o}ckmann et al.}{2008}]{loc08} L\"{o}ckmann U., Baumgardt H., Kroupa P., 2008, ApJ Lett., 683, L151
\bibitem[\protect\citeauthoryear{Marks \& Kroupa}{2010}]{mar10} Marks M., Kroupa P., 2010, MNRAS, 406, 200
\bibitem[\protect\citeauthoryear{Marks \& Kroupa}{2012}]{mar12} Marks M., Kroupa P., 2012, A\& A, 543, A8
\bibitem[\protect\citeauthoryear{Marks, Kroupa \& Baumgardt}{Marks et al.}{2008}]{mar08} Marks M., Kroupa P., Baumgardt H., 2008, MNRAS, 386, 2047
\bibitem[\protect\citeauthoryear{Marks et al.}{2012}]{mkd12} Marks M., Kroupa P., Dabringhausen J., Pawlowski M.S., 2012, MNRAS, 422, 2246
\bibitem[\protect\citeauthoryear{Miyamoto \& Nagai}{1975}]{miy75} Miyamoto M., Nagai R., 1975, Publ. Astron. Soc. Jpn., 27, 533
\bibitem[\protect\citeauthoryear{Navarro, Frenk \& White}{Navarro et al.}{1997}]{nav97} Navarro J.F., Frenk C.S., White S.D., 1997, ApJ, 490, 493
\bibitem[\protect\citeauthoryear{N\'{e}meth et al.}{2016}]{nem16} N\'{e}meth P., et al., 2016, ApJ Lett., 821, L13
\bibitem[\protect\citeauthoryear{O'Leary \& Loeb}{2008}]{oll08} O'Leary R. M., Loeb A., 2008, MNRAS, 383, 86
\bibitem[\protect\citeauthoryear{Oh, Kroupa \& Pflamm-Altenburg}{Oh et al.}{2015}]{ohk15} Oh S., Kroupa P., Pflamm-Altenburg J., 2015, ApJ, 805, 92
\bibitem[\protect\citeauthoryear{Pawlowski, Kroupa \& Jerjen}{Pawlowski et al.}{2013}]{paw13} Pawlowski M.S., Kroupa P., Jerjen H., 2013, MNRAS, 435, 1928
\bibitem[\protect\citeauthoryear{Pawlowski, Pflamm-Altenburg \& Kroupa}{Pawlowski et al.}{2012}]{paw12} Pawlowski M.S., Pflamm-Altenburg J., Kroupa P., 2012, MNRAS, 423, 1109
\bibitem[\protect\citeauthoryear{Perets, Hopman \& Alexander}{Perets et al.}{2007}]{per07} Perets H.B., Hopman C., Alexander T., 2007, ApJ, 656, 709
\bibitem[\protect\citeauthoryear{Perets}{2009}]{per09} Perets H.B., 2009, ApJ, 698, 1330
\bibitem[\protect\citeauthoryear{Perets et al.}{2009}]{pee09} Perets H.B. et al. 2009, ApJ, 697.2, 2096
\bibitem[\protect\citeauthoryear{Pflamm-Altenburg, Weidner \& Kroupa}{Pflamm-Altenburg et al.}{2007}]{pfl07} Pflamm-Altenburg J., Weidner C., Kroupa P., 2007, ApJ, 671, 1550
\bibitem[\protect\citeauthoryear{Plummer}{1911}]{plu11} Plummer H.C., 1911, MNRAS, 71, 140
\bibitem[\protect\citeauthoryear{Reid et al.}{2014}]{rei14} Reid M.J. et al., 2014, ApJ, 783, 130
\bibitem[\protect\citeauthoryear{Sana et al.}{2012}]{san12} Sana H. et al., 2012, Science, 337, 444
\bibitem[\protect\citeauthoryear{Sari, Kobayashi \& Rossi}{Sari et al.}{2009}]{sar09} Sari R., Kobayashi S., Rossi E.M., 2010, ApJ, 708, 605
\bibitem[\protect\citeauthoryear{Sesana, Haardt \& Madau}{Sesana et al.}{2007}]{ses07} Sesana A., Haardt F., Madau P., 2007, MNRAS Lett., 379, L45
\bibitem[\protect\citeauthoryear{Sherwin, Loeb \& O'Leary}{Sherwin et al.}{2008}]{she08} Sherwin B., Loeb A., O'Leary R., 2008, MNRAS, 386, 1179
\bibitem[\protect\citeauthoryear{Silva \& Napiwotzki}{2011}]{sil11} Silva M.D.V., Napiwotzki R., 2011, MNRAS, 411, 2596
\bibitem[\protect\citeauthoryear{\v{S}ubr \& Haas}{2014}]{sub14} \v{S}ubr L., Haas J., 2014, ApJ, 786, 121
\bibitem[\protect\citeauthoryear{\v{S}ubr \& Haas}{2016}]{sub16} \v{S}ubr L., Haas J., 2016, ApJ, 828, 1
\bibitem[\protect\citeauthoryear{Tauris}{2015}]{tau15} Tauris T.M., 2015, MNRAS Lett., 448, L6
\bibitem[\protect\citeauthoryear{Weidner \& Kroupa}{2004}]{wei04} Weidner C., Kroupa P., 2004, MNRAS, 348, 187
\bibitem[\protect\citeauthoryear{Yu \& Madau}{2007}]{yum07} Yu Q., Madau P., 2007, MNRAS, 379, 1293
\bibitem[\protect\citeauthoryear{Yu \& Tremaine}{2003}]{yut03} Yu Q., Tremaine S., 2003, ApJ, 599, 1129
\bibitem[\protect\citeauthoryear{Ziegerer et al.}{2015}]{zie15} Ziegerer E., Volkert M., Heber U., Irrgang A., G\"{a}nsicke B.T., Geier S., 2015, A\& A, 576, L14
\bibitem[\protect\citeauthoryear{Zubovas, Wynn \& Gualandris}{Zubovas et al.}{2013}]{zub13} Zubovas K., Wynn G.A., Gualandris A., 2013, ApJ, 771.2, 118
\end{thebibliography}
\end{document}